\documentclass[aip,graphicx]{revtex4-1}
\draft 

\usepackage{graphicx}        
\usepackage{natbib}         
\usepackage{amssymb}        
\usepackage{amsmath}
\usepackage{color}           
\usepackage{url}             
\usepackage{epstopdf}
\usepackage{soul}

\providecommand\bnabla{\boldsymbol{\nabla}}
\providecommand\bcdot{\boldsymbol{\cdot}}
\newcommand\bu{{\boldsymbol{u}}}

\newcommand\be{{\boldsymbol{e}}}
\newcommand\bA{{\boldsymbol{A}}}
\newcommand\bB{{\boldsymbol{B}}}
\newcommand\bF{{\boldsymbol{F}}}
\newcommand\bx{{\boldsymbol{x}}}
\newcommand\by{{\boldsymbol{y}}}
\newcommand\bz{{\boldsymbol{z}}}

\renewcommand\Re{{\rm Re}}
\newcommand\Ri{{\rm Ri}}
\newcommand\Pe{{\rm Pe}}
\newcommand\p{\partial}
\newcommand\md{\mathrm{d}}

\begin{document}

\pagestyle{empty} 


\title{The stability of stratified spatially periodic shear flows at low P\'eclet number}

\author{Pascale Garaud}
\email[]{pgaraud@ucsc.edu} 
\affiliation{Department of Applied Mathematics and Statistics, Baskin School of Engineering, University of California at Santa Cruz, 1156 High Street, Santa Cruz CA 95064, USA}
\author{Basile Gallet} 
\affiliation{Service de Physique de l'Etat Condens\'e, DSM/IRAMIS, CNRS UMR 3680, CEA Saclay, 91191 Gif-sur-Yvette cedex, France. }
\author{Tobias Bischoff}
\affiliation{Division of Geological and Planetary Sciences, California Institute of Technology, Mail Code 170-25, 1200 E. California Blvd., Pasadena, CA 91125, USA}
 
\date{\today}

\begin{abstract}
This work addresses the question of the stability of stratified, spatially periodic shear flows at low P\'eclet number but high Reynolds number. This little-studied limit is motivated by astrophysical systems, where the Prandtl number is often very small. Furthermore, it can be studied using a reduced set of ``low-P\'eclet-number equations'' proposed by Lignieres [Astronomy \& Astrophysics, 348, 933-939, 1999]. Through a linear stability analysis, we first determine the conditions for instability to infinitesimal perturbations. We formally extend Squire's theorem to the low-P\'eclet-number equations, which shows that the first unstable mode is always two-dimensional. We then perform an energy stability analysis of the low-P\'eclet-number equations and prove that for a given value of the Reynolds number, above a critical strength of the stratification, {\it any} smooth periodic shear flow is stable to perturbations of arbitrary amplitude. In that parameter regime, the flow can only be laminar and turbulent mixing does not take place. Finding that the conditions for linear and energy stability are different, we thus identify a region in parameter space where finite-amplitude instabilities could exist. Using direct numerical simulations, we indeed find that the system is subject to such finite-amplitude instabilities. We determine numerically how far into the linearly stable region of parameter space turbulence can be sustained. 
\end{abstract}

\maketitle


\section{Introduction}
\label{sec:intro}

The study of the onset of turbulence in stratified shear flows has a long history that dates back to \citet{Richardson1920}. He argued that the kinetic energy of turbulent eddies in a stratified shear flow 
can only decrease if $N^2/S^2 > 1$, where $N$ is the local buoyancy frequency, and $S = |d \bu/dz|$ is the local shearing rate of the flow field $\bu$ in the vertical direction $\be_z$. This criterion, derived simply from energetic arguments, is now commonly referred to as Richardson's criterion, and the local ratio 
\begin{equation}
J(z) = \frac{N^2(z)}{S^2(z)} 
\end{equation}
is called the gradient Richardson number. The first linear stability analysis of a stratified shear flow is due to \citet{Taylor1931}, who considered both continuously and discretely varying stratification and shear profiles. This work, together with \citet{Goldstein1931}, then led to the derivation of the Taylor-Goldstein eigenvalue equation for the complex growth rate of two-dimensional infinitesimal disturbances in stratified shear flows. The solution of this equation for a given shear profile $S(z)$ and stratification profile $N(z)$ can be obtained either analytically in a few particular cases, or numerically in general. It wasn't until much later, however, that the first general result on the stability of stratified shear flows was derived by \citet{Miles61} and \citet{Howard61}: a system is stable to infinitesimal perturbations provided $J(z)$ is everywhere larger than 1/4. As discussed by \citet{HowardMaslowe1973}, this theorem should not be viewed as a refinement of Richardson's argument (i.e. replacing 1 by 1/4), since the latter was specifically interested in determining when turbulence could be sustained, rather than triggered. In this sense, Richardson's original argument should be viewed more as a nonlinear stability criterion than a linear one. 

These results were obtained in the limit of vanishing viscosity and diffusivity. For thermally stratified flows, however, thermal diffusion can have a significant influence on the development of shear instabilities by damping the buoyancy restoring force. This effect was first studied by \citet{Townsend58} in the context of atmospheric flows. He showed that the thermal adjustment of the fluid parcel to its surroundings, by radiative heating and cooling or by thermal conduction, always acts to destabilize the flow and increases the critical Richardson number for linear stability, $J_{\rm crit}$, by a factor inversely proportional to the product of the shearing rate $S$ with the cooling time $t_{\rm cool}$ (this product is a local P\'eclet number for the flow), so $J_{\rm crit} \sim (t_{\rm cool} S)^{-1}$. Viscosity, meanwhile, has a generally stabilizing influence \citep{Dudis1974}. \citet{Zahn1974} emphasized the importance of these results for stellar astrophysics: in stellar interiors where the Prandtl number is typically very small ($\Pr \sim 10^{-8} - 10^{-5}$)  high Reynolds number flows can also have a low P\'eclet number, or in other words, thermally diffusive shear flows exist when viscosity is nevertheless small enough not to suppress the development of the instability. This combination is ideal for shear instabilities, and is specific to astrophysical systems -- it cannot happen for most geophysical flows where the Prandtl number is usually of order unity or larger. 

Applying \citet{Townsend58}'s results to shear-induced turbulence in stellar interiors, \citet{Zahn1974} further argued that the relevant cooling timescale is the radiative timescale based on the size $l$ of turbulent eddies, namely $t_{\rm cool} = l^2 / \kappa_T$ where $\kappa_T$ is the thermal diffusivity. He then proposed to take for $l$ the smallest length scale for which viscosity is still negligible, that is, one for which the turbulent Reynolds number $\Re_l = S l^2 / \nu = S t_{\rm cool}/ \Pr \sim \Re_{\rm crit}$ (where $\nu$ is the kinematic viscosity), where $\Re_{\rm crit}$ is a constant that he estimates to be around $10^3$. This would imply $J_{\rm crit} \sim (S t_{\rm cool})^{-1} \sim \Re^{-1}_{\rm crit} \Pr^{-1}$, or in other words, $J_{\rm crit} \Pr \sim \Re^{-1}_{\rm crit} \sim 10^{-3}$. \citet{Zahn1974}'s argument, as in the case of \citet{Richardson1920}'s original argument, should be viewed as a nonlinear stability criterion rather than a linear one, since it relies on the presence of pre-existing turbulent eddies.

Zahn's work had an enormous impact in the field of stellar evolution. While the standard Richardson criterion is far too stringent to allow for the development of shear instabilities in the absence of thermal diffusion (the typical Richardson number being much larger than one even in the strongest known stellar shear layers), its relaxation allows for the possibility of much-needed mixing in stellar evolution theory. Indeed, models without any form of turbulent mixing in stably stratified regions are not able to account for observations. As reviewed by \citet{Pinsonneault97} the problem is particularly acute when it comes to explaining the surface chemical abundances of light elements such as lithium and beryllium, as well as products and by-products of nuclear reactions such as helium, carbon, nitrogen and oxygen. More recently, further indication of the need for turbulent mixing was revealed by asteroseisomolgy thanks to the Kepler mission. Measurements of the internal rotation rate of red giant stars \citep{Mosser12b} are inconsistent with evolution models in which turbulent angular-momentum transport in stably stratified regions is neglected. In both cases, therefore, efficient chemical transport and angular-momentum transport by shear instabilities could be the key to resolving these problems -- the question remains, however, of whether these instabilities are indeed triggered, and how efficient mixing is. 

In the limit of low-P\'eclet numbers (i.e., high thermal diffusivity), the temperature fluctuations are slaved to the vertical velocity. The corresponding quasi-static approximation was originally introduced to study low-Prandtl-number thermal convection\citep{Spiegel1962, Thual}. In the context of stably-stratified systems, the quasi-static approximation was introduced only recently by \citet{Lignieres1999} (see Section \ref{sec:lowPeeqs} for more detail). He showed that the standard Boussinesq equations can be replaced by a reduced model that is valid in the asymptotic low-P\'eclet-number limit, and that this model only depends on two parameters: the Reynolds number, and the product of the Richardson number with the P\'eclet number. As a result, the linear stability properties of the system depend on the product $\Pe_SJ$ (where $\Pe_S = S L_S^2 / \kappa_T$ with $L_S$ being a characteristic vertical length scale of the laminar flow) rather than on each parameter individually. Since $\Pe_S$ is small by assumption, shear-induced turbulence can be expected even if the Richardson number is much larger than one. 

By contrast with linear theory, very little is known to date about the stability of stratified shear flows to finite-amplitude perturbations when viscosity and thermal diffusivity are both taken into account. 
It is yet a question of crucial importance in stellar astrophysics, since the presence or absence of vertical mixing can strongly affect model predictions. We therefore set out in this work to characterize the domain of instability of low-P\'eclet-number shear flows to finite-amplitude perturbations. For simplicity, we consider a specific shear profile that is periodic in the vertical direction.  We present the model in Section \ref{sec:model}, and briefly discuss the low-P\'eclet-number asymptotic equations proposed by \citet{Lignieres1999}. In Sections \ref{sec:LS} and \ref{sec:ES}, we study the linear and nonlinear stability of the system respectively, and contrast the results in the low-P\'eclet-number approximation to those obtained starting from the full set of primitive equations. As we shall demonstrate using an energy stability analysis of the low-P\'eclet number equations, smooth periodic shear flows are stable to perturbations of arbitrary amplitude for sufficiently large Richardson number, for a given value of the Reynolds number. In Section \ref{sec:num}, we turn to direct numerical simulations to study the transition to turbulence via linear instabilities and finite-amplitude instabilities. 
We summarize our results and conclude in Section \ref{sec:ccl}.

\section{The model}
\label{sec:model}

\subsection{Model setup}

\begin{figure}[h]
\centerline{\includegraphics[width=0.75\textwidth]{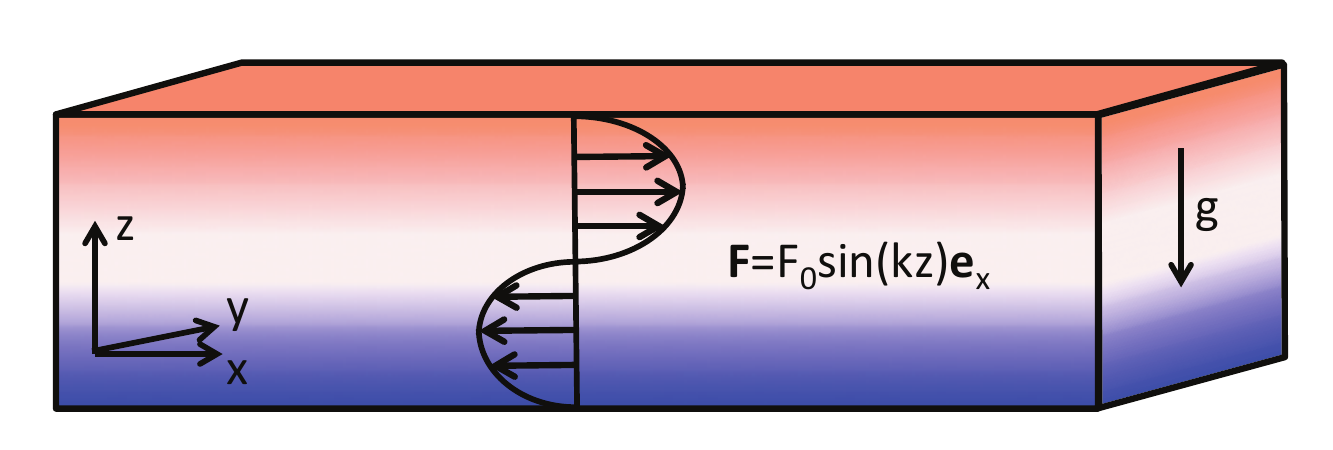}}
\caption{Model setup: a horizontal shear flow is driven by a body-force. The background stratification is linear, and the temperature and velocity fluctuations are periodic in the three directions.}
\label{fig:Schematic}
\end{figure}

Since our intention is to study the energy stability properties of stratified shear flows, it is crucial to start with a model where the mechanism driving the shear is explicit, which guarantees a well-defined energy budget. Two options are available: boundary-forcing, and body-forcing. Having potential applications to stellar astrophysics in mind, we prefer the latter in order to avoid boundary layer dynamics near solid walls, which are rarely present in stars.

A simple and numerically-efficient way of studying body-forced, stratified shear flows is to 
consider a Boussinesq system \citep{SpiegelVeronis1960}, where the forcing and all the perturbations are triply-periodic, and where the background density is linearly stratified\footnote{We ignore in this paper the effects of compositional stratification.} (see Figure \ref{fig:Schematic}). The model equations describing such a system are:
\begin{eqnarray}
&& \frac{\p \bu}{\p t} + \bu \bcdot \bnabla \bu = - \frac{1}{\rho_0} \bnabla p + \alpha g T \be_z + \nu \nabla^2 \bu + \frac{1}{\rho_0} \bF  \mbox{  ,}\label{eq:original1}\\
&& \bnabla \bcdot \bu = 0\mbox{  ,} \\
&& \frac{\p T}{\p t} + \bu \cdot \bnabla T + w T_{0z} = \kappa_T \nabla^2 T\mbox{  ,}
\label{eq:original}
\end{eqnarray}
where $\bu = (u,v,w)$ is the triply-periodic velocity field, $p$ and $T$ are the triply-periodic pressure and temperature perturbations, $\rho_0$ is the mean density of the region considered, 
$\alpha$ is the coefficient of thermal expansion, $g$ is gravity, $\nu$ and $\kappa_T$ are the viscosity and thermal diffusivity (respectively). The quantities 
$\rho_0$, $\alpha$, $g$, $\nu$, $\kappa$ are all assumed to be constant, as in the standard Boussinesq approximation. The use of the latter is justified as long as the vertical height of the domain is much smaller than a density scaleheight. Finally, we assume that there is a constant background temperature gradient\footnote{To be precise, since the fluid is weakly compressible in stellar interiors, the background temperature gradient $T_{0z}$ should be replaced by the potential temperature gradient, i.e. the difference between the actual temperature gradient and the local adiabatic temperature gradient, as shown by \citet{SpiegelVeronis1960}.}  $T_{0z}$, and that all thermodynamical and dynamical perturbations have zero mean in the domain. 

The applied force $\bF$ should be triply-periodic as well. A natural candidate is  a sinusoidal forcing, thus driving a Kolmogorov flow in the laminar regime. 
In what follows, we assume that $\bF$ is of the form
\begin{equation}
\bF = F_0 \sin(kz) \be_x\mbox{  ,}
\end{equation}
which defines a typical lengthscale $k^{-1}$. In the steady laminar regime, this force generates a sinusoidal shear flow along the $x$-direction,
\begin{equation}
\bu_L= \frac{F_0}{\rho_0 \nu k^2} \sin(kz) \, {\bf e}_x \, . \label{lamsoldimensional}
\end{equation}
Note that while the present paper deals mostly with Kolmogorov forcing, the energy stability of arbitrary smooth velocity profiles is discussed in section \ref{sec:upperbound}.


\subsection{Non-dimensionalization and model parameters}
\label{sec:nondim}

We non-dimensionalize the equations using the amplitude of the laminar solution, $\frac{F_0}{\rho_0 \nu k^2}$, as a velocity scale. We also use the spatial scale of the laminar solution, $k^{-1}$, as the unit length scale. This then defines the timescale $\frac{k \rho_0 \nu}{F_0}$. With this choice of units, the equations (\ref{eq:original1})-(\ref{eq:original}) become
\begin{eqnarray}
&& \frac{\p \bu}{\p t} +  \bu \bcdot \bnabla \bu = -  \bnabla  p + \Ri \, T \be_z + \frac{1}{\Re} \nabla^2 (\bu- \sin(z) \be_x)  \mbox{  ,} \label{eq:originallaminar1}\\
&& \bnabla \bcdot  \bu = 0 \mbox{  ,}\\
&& \frac{\p T}{\p t} + \bu \cdot \bnabla T + w = \frac{1}{\Pe} \nabla^2 T\mbox{  ,}
\label{eq:originallaminar}
\end{eqnarray}
where 
\begin{equation}
\Re = \frac{F_0}{\rho_0 \nu^2 k^3} \, , \quad  \Ri = \frac{\alpha g T_{0z} \rho_0^2 \nu^2 k^2}{F_0^2}   \, , \quad \Pe = \frac{F_0}{\rho_0 \nu k^3 \kappa_T} \, .\label{param}
\end{equation} 
The laminar solution (\ref{lamsoldimensional}) now takes the dimensionless form $\bu_L=\sin(z) \, \be_x$.
Provided the system remains in the vicinity of this laminar solution, $\Re$, $\Pe$ and $\Ri$ are the usual Reynolds, P\'eclet and Richardson numbers based on the typical velocity of the flow.

\subsection{Low P\'eclet number approximation}
\label{sec:lowPeeqs}

When a field diffuses on a timescale much shorter than the advective time, it enters a {\it quasi-static} regime where the source term and diffusive term instantaneously balance. Such a quasi-static regime has been used for decades in the context of magneto-hydrodynamics of liquid metals: at low magnetic Reynolds number, the induced magnetic field is slaved to the velocity field \citep[see for instance][]{Sommeria82}. The equivalent approximation for flows of low Prandtl number fluids was originally introduced  in the context of thermal convection \citep{Spiegel1962,Thual}. Rather surprisingly, this quasi-static approximation appeared only much more recently in the context of stably-stratified flows. 

\citet{Lignieres1999} proposed that in the low-P\'eclet-number limit the governing equations (\ref{eq:originallaminar1})-(\ref{eq:originallaminar}) can be approximated by the reduced set of so-called ``low-P\'eclet-number" equations (LPN equations hereafter)
\begin{eqnarray}
&& \frac{\p \bu}{\p t} + \bu \bcdot \bnabla \bu = - \bnabla  p + \Ri  T \be_z + \frac{1}{\Re}  \nabla^2  \bu +  \bF  \mbox{  ,} \\
&& \bnabla \bcdot  \bu = 0  \mbox{  ,} \\
&&  w = \frac{1}{\Pe} \nabla^2  T  \mbox{  ,}
\label{eq:lowPeT}
\end{eqnarray}
to zeroth and first order
 in $\Pe$. To derive equation (\ref{eq:lowPeT}), one can assume a regular expansion of $T$ in $\Pe$, as in $T = T_0 + \Pe T_1 + O(\Pe^2)$, and further assume that the velocity field is of order unity. At the lowest order, the temperature equation yields $\nabla^2  T_0 = 0$ which then implies $T_0 = 0$ given the applied boundary conditions. The next order then yields the quasi-static balance $w = \nabla^2 T_1 \simeq \Pe^{-1} \nabla^2 T$ as required. It is worth mentioning that in \citet{Lignieres1999}'s original work the velocity is scaled with its dimensional r.m.s. value $u_{\rm rms}$, so it is $\Pe_{\rm rms} = u_{\rm rms}/ k \kappa_T$, rather than $\Pe$, that has to be small for the LPN equations to be valid. In Sections \ref{sec:LS} and \ref{sec:ES}, we shall study the linear and energy stability of a laminar flow for which the r.m.s. velocity is of the same order as the flow amplitude. In that case $\Pe \simeq \Pe_{\rm rms}$ and we expect the LPN equations to be valid whenever $\Pe$ is small. In Section \ref{sec:num}, however, we shall see that numerical simulations of turbulent shear flows that have large $\Pe$ can still be well-described by the LPN equations as long as $\Pe_{\rm rms}$ is small.
  
It is also worth noting that the LPN equations only allow for temperature fluctuations $T$ that have a zero horizontal mean, by contrast with the full equations. As a result, the horizontal mean of the full temperature field (background plus perturbations) is necessarily linear in $z$. To see this, we take the horizontal average of the thermal equation, which, assuming that there is no vertical mean flow (which can be guaranteed by making sure the initial conditions do not have one), results in 
 \begin{equation}
 \frac{\partial^2 \langle T \rangle_h }{\partial z^2} = 0 \mbox{  ,}
 \end{equation}
in the LPN equations, where  $ \langle T \rangle_h$ is the horizontal average of $T$. The only solution of this equation which satisfies periodicity is the constant solution; further requiring that the volume-average of $T$ be zero then implies that $ \langle T \rangle_h = 0$. By contrast, taking the horizontal average of the standard temperature equation under the same assumptions would result in 
 \begin{equation}
\frac{\partial  \langle T \rangle_h}{\partial t} +  \frac{\partial}{\partial z} \langle wT \rangle_h = \frac{1}{\Pe}  \frac{\partial^2 \langle T \rangle_h }{\partial z^2} \mbox{  ,}
 \end{equation}
which has solutions with non-zero $ \langle T \rangle_h$. These solutions can, for instance, develop into density staircases under the right circumstances\citep{Phillips1972,Balmforthal1998}. The latter are however prohibited in the LPN equations. Among other effects, this rules out the development of Holmboe modes\citep{Holmboe62}, and may explain why it is possible to get simple energy stability results in the LPN limit but not for the full equations.

Finally, combining the momentum and the thermal energy equations yields
\begin{equation}
\frac{\p \bu}{\p t} + \bu \bcdot \bnabla \bu = - \bnabla  p + \Ri \Pe \nabla^{-2} w \be_z + \frac{1}{\Re}  \nabla^2  \bu +  \bF  \mbox{  ,}
\label{eq:lowPeeqs}
\end{equation}
which formally shows that the Richardson number is no longer the relevant parameter of the system, but that $\Ri \Pe$ is. The work of \citet{Lignieres1999} thus puts the arguments of \citet{Townsend58} and \citet{Zahn1974} discussed in Section \ref{sec:intro} on a firm theoretical footing. \citet{Lignieres1999} and \citet{Lignieresetal1999} verified that the LPN equations correctly account for the linear stability properties of various systems in the low-P\'eclet-number limit. \citet{PratLignieres13} later also verified that they  correctly reproduced the low-P\'eclet dynamics of their 3D nonlinear simulations. In this paper, we continue to verify the validity of the LPN equations through stability analyses and nonlinear simulations. 
 
\section{Linear stability of a periodic Kolmogorov flow}
\label{sec:LS} 

%

We first focus on the stability of the laminar solution to infinitesimal perturbations. We solve the linearized versions of equations (\ref{eq:originallaminar1})-(\ref{eq:originallaminar}),
%
\begin{eqnarray}
&& \frac{\p \bu'}{\p t} +  \bu_L \bcdot \bnabla \bu' +  \bu' \bcdot \bnabla \bu_L =  -  \bnabla  p + \Ri T' \be_z + \frac{1}{\Re} \nabla^2 \bu'  \mbox{  ,}  \label{eq:linear21}\\
&& \bnabla \bcdot  \bu' = 0 \mbox{  ,} \\
&& \frac{\p T'}{\p t} + \bu_L \cdot \bnabla T' + w' = \frac{1}{\Pe} \nabla^2 T'\mbox{  ,}
\label{eq:linear2}
\end{eqnarray}
where $\bu'$ and $T'$ are infinitesimal perturbations to the linearly stratified background shear flow $\bu_L = \sin(z) \be_x$. 

\subsection{Squire's transformation}

The linear stability of the unstratified Kolmogorov flow $\bu_L$ was first investigated in detail by \citet{Beaumont81}. Squire's theorem \citep{Squire33} states that the first unstable mode as the Reynolds number increases is a ($y$-independent) 2D mode. This strong result implies that one can focus on 2D perturbations to determine the stability threshold of the system. Such a 2D analysis is much simpler and computationally less expensive than a 3D one.
\citet{Beaumont81} found that 2D flows are unstable only for $\Re \ge \sqrt{2}$.

The linear stability of the stratified Kolmogorov flow $\bu_L$ to 2D perturbations was studied in detail by \citet{BalmforthYoung2002}. The 2D case can be made more generally relevant by noting that Squire's transformation \citep{Squire33} for the viscous unstratified case can be extended to the stratified case with thermal diffusion to argue that the linear stability of any 3D mode can equivalently be studied by considering that of a 2D mode at lower or equal Reynolds and P\'eclet numbers, and higher or equal Richardson number. This result, which was summarily discussed by \citet{Yih55} and clarified by \citet{Smythetal88} and \citet{SmythPeltier90} \citep[see also][]{DrazinReid}, states that the growth rate $\lambda_3$ of the 3D normal mode $q_3(x,y,z,t) = \hat{q}_3(z) \exp(ilx + imy + \lambda_3 t)$ at parameters $(\Re,\Pe,\Ri)$ is related to that of the 2D normal mode $q_2(x,y,z,t) = \hat{q}_2(z) \exp(iLx  + \lambda_2 t)$ at suitably rescaled parameters via
\begin{equation}
\lambda_3 \equiv f(l,m;\Re,\Pe,\Ri) = \frac{l}{L} \lambda_2 \equiv \frac{l}{L} f\left(L,0;  \frac{l}{L}\Re,  \frac{l}{L}\Pe , \frac{L^2}{l^2} \Ri   \right) \mbox{  ,}
\label{eq:l2l3}
\end{equation}
where $L = \sqrt{l^2 + m^2}$. One can apply the same method to the LPN equations, and the result can readily be deduced from (\ref{eq:l2l3}). Indeed, the transformation (\ref{eq:l2l3}) is valid for any P\'eclet number, so it remains valid for low P\'eclet numbers. In this limit, we saw that only the product $\Ri \Pe$ is a relevant parameter: as a consequence, one can replace the last two arguments of the function $f$ in (\ref{eq:l2l3}) by the product of the two. The low-P\'eclet version of Squire's transformation therefore gives  
\begin{equation}
\lambda_3 \equiv f(l,m;\Re,\Ri\Pe) = \frac{l}{L} \lambda_2 \equiv \frac{l}{L} f\left(L,0;  \frac{l}{L}\Re, \frac{L}{l} \Ri\Pe   \right) \mbox{  .}
\label{eq:l2l3lowPe}
\end{equation}

This relationship between the growth rates of 2D and 3D modes has important implications for the marginal linear stability surface. In order to find the latter in 2D, we first maximize the real part of $f(L,0;\Re,\Pe,\Ri)$ over all possible values of $L$, yielding the function $S_2(\Re,\Pe,\Ri)$ which returns the growth rate of the fastest growing mode for each parameter set $(\Re,\Pe,\Ri)$. The marginal linear stability surface is then defined by $S_2 = 0$. Similarly, the marginal linear stability surface for 3D perturbations is obtained by constructing the function $S_3(\Re,\Pe,\Ri) = \max_{l,m} Re [ f(l,m;\Re,\Pe,\Ri) ] $ and setting $S_3 = 0$.  If the functions $S_2$ and $S_3$ are the same, then so are the surfaces $S_2 = 0$ and $S_3 = 0$, which implies in turn that the first modes to be destabilized are the 2D modes. This is the case for instance in the limit where stratification is negligible (see above). 

In general, the only way to determine whether $S_2 = S_3$ for a given linear stability problem is to construct these functions by brute force, using their original definition as the growth rates of the fastest growing modes. While this is not too time-consuming in 2D, it can become computationally expensive in 3D. However in this particular problem, since the growth rates of 2D and 3D modes are related, we also have: 
\begin{equation}
S_3(\Re,\Pe,\Ri) = \max_{\chi\in [0,1]} \chi S_2(\chi \Re,\chi \Pe,\Ri / \chi^2)\mbox{  ,}
\label{eq:s3s2}
\end{equation}
where $\chi = | l / L|$. A similar relationship applies for the LPN equations: 
\begin{equation}
S_3(\Re,\Ri\Pe) = \max_{\chi\in [0,1]} \chi S_2(\chi \Re,\Ri\Pe / \chi)\mbox{  .}
\label{eq:s3s2lowPe}
\end{equation}
Note that it is easier to construct $S_3$ from equations (\ref{eq:s3s2}) or (\ref{eq:s3s2lowPe}) than to do so directly. 

Whether $S_2 = S_3$ or not then simply depends on the properties of $S_2$. It is quite easy to find sufficient conditions that guarantee $S_2 = S_3$.  For instance, in the case of the standard equations, if $S_2$ is a strictly increasing function of both $\Re$ and $\Pe$, and a strictly decreasing function of $\Ri$, then the maximum over all possible values of $\chi$ is achieved for $\chi = 1$, which ensures that $S_2 = S_3$. For the LPN equations, it is sufficient to show that $S_2$ is a strictly increasing function of $\Re$ and a strictly decreasing function of $\Ri\Pe$.
In what follows, we therefore first study the stability of 2D modes, and then use these results to conclude on the stability of the system to 3D modes.


\subsection{Linear stability analysis using Floquet theory} 


We use a stream function to describe divergence-free 2D perturbations,
\begin{equation}
\bu = \bu_L + \nabla \times \left(\psi' \be_y \right) \mbox{  ,}
\end{equation}
where $\psi'$ is the infinitesimal perturbation.  The linearized equations (\ref{eq:linear21})-(\ref{eq:linear2}) become
\begin{eqnarray}
\frac{\p }{\p t}(\nabla^2  \psi') + \sin(z)   \left( \frac{\p}{\p x}  (\nabla^2 \psi' +  \psi') \right) = \Ri  \frac{\p  T'}{\p x}  + \frac{1}{\Re}  \nabla^4   \psi'  \mbox{  ,} \nonumber \\
\frac{\p T'}{\p t}  + \sin(z) \frac{\p T'}{\p x} + \frac{\p  \psi'}{\p x} =  \frac{1}{\Pe}  \nabla^2  T' \mbox{  .}
\end{eqnarray}
This set of PDEs for $ T'$ and $\psi'$ has coefficients that are independent of $t$ and $x$, but periodic in $z$. Normal modes for this system are of the form
\begin{equation}
q'(x,z,t) = e^{iLx + \lambda t} \hat q(z)  \mbox{  ,}
\end{equation}
where $q'$ is either $T'$ or $\psi'$, and $L$ is real. Using Floquet theory, we then seek solutions for $\hat q$ given by
\begin{equation}
\hat q(z) =  e^{iaz} \sum_{n=-N}^{N} q_n e^{inz}  \mbox{  ,}
\end{equation}
where $a$ is real, to satisfy the general periodicity of the system. Substituting this ansatz into the previous equations, we obtain an algebraic system for the $\psi_n$ and $T_n$:
 \begin{eqnarray}
&& - \lambda ((a+n)^2 + L^2) \psi_n + \frac{L}{2} \left[ (1 - (a+n-1)^2- L^2) \psi_{n-1} -(1 - (a+n+1)^2- L^2) \psi_{n+1}   \right] \nonumber \\ 
&& = i \Ri L T_n + \frac{1}{\Re} ((a+n)^2 + L^2)^2 \psi_n  \mbox{  ,} \nonumber \\ 
&& \lambda T_n   + \frac{L}{2} \left[ T_{n-1} - T_{n+1} \right]   + iL \psi_n  =  - \frac{1}{\Pe} ((a+n)^2 + L^2) T_n  \mbox{  ,}
\end{eqnarray}
for $n = -N ... N$. This can be cast as the linear eigenproblem, 
\begin{equation}
{\bf M}(L;a;\Re,\Pe,\Ri) \bx = \lambda  \bx \mbox{  ,}
\end{equation}
where $\bx = \{ \psi_{-N}, ... , \psi_N, T_{-N}, ... , T_N\}$,  which can be solved for the complex growth rate $\lambda$. The real part of the latter can then be maximized over all possible values of $a$ and $L$ for given system parameters $(\Re, \Pe,\Ri)$ to determine the temporal behavior and spatial structure of the most rapidly growing mode of the shear instability, or in other words, to construct $S_2$ (see previous section). In both unstratified and stratified cases studied so far, the first unstable modes at the instability threshold have the same periodicity in $z$ as that of the background shear, so that $a=0$ \citep{Beaumont81,Gotoh83,BalmforthYoung2002}. We verified that this is indeed the case here as well. In what follows, we therefore restrict the presentation of our results to the case $a=0$. 

The equivalent problem for the LPN equations is given by 
  \begin{eqnarray}
&& - \lambda ((a+n)^2 + L^2) \psi_n + \frac{L}{2} \left[ (1 - (a+n-1)^2- L^2) \psi_{n-1} -(1 - (a+n+1)^2- L^2) \psi_{n+1}   \right] \nonumber \\ 
&& =  \frac{L^2  }{ (a+n)^2 + L^2 } \Ri \Pe   \psi_n + \frac{1}{\Re} ((a+n)^2 + L^2)^2 \psi_n  \mbox{  ,}
\end{eqnarray}
which can be cast as
\begin{equation}
{\bf M'}(L;a;\Re, \Ri\Pe) \by = \lambda \by \mbox{  ,}
\end{equation}
where $\by = \{ \psi_{-N}, ... , \psi_N\}$. This time, the fastest growing mode only depends on two system parameters, namely $\Re$ and the product $\Ri\Pe$. Again, we restrict the following analysis to the case $a=0$.

\begin{figure}[h]
\centerline{\includegraphics[width=0.9\textwidth]{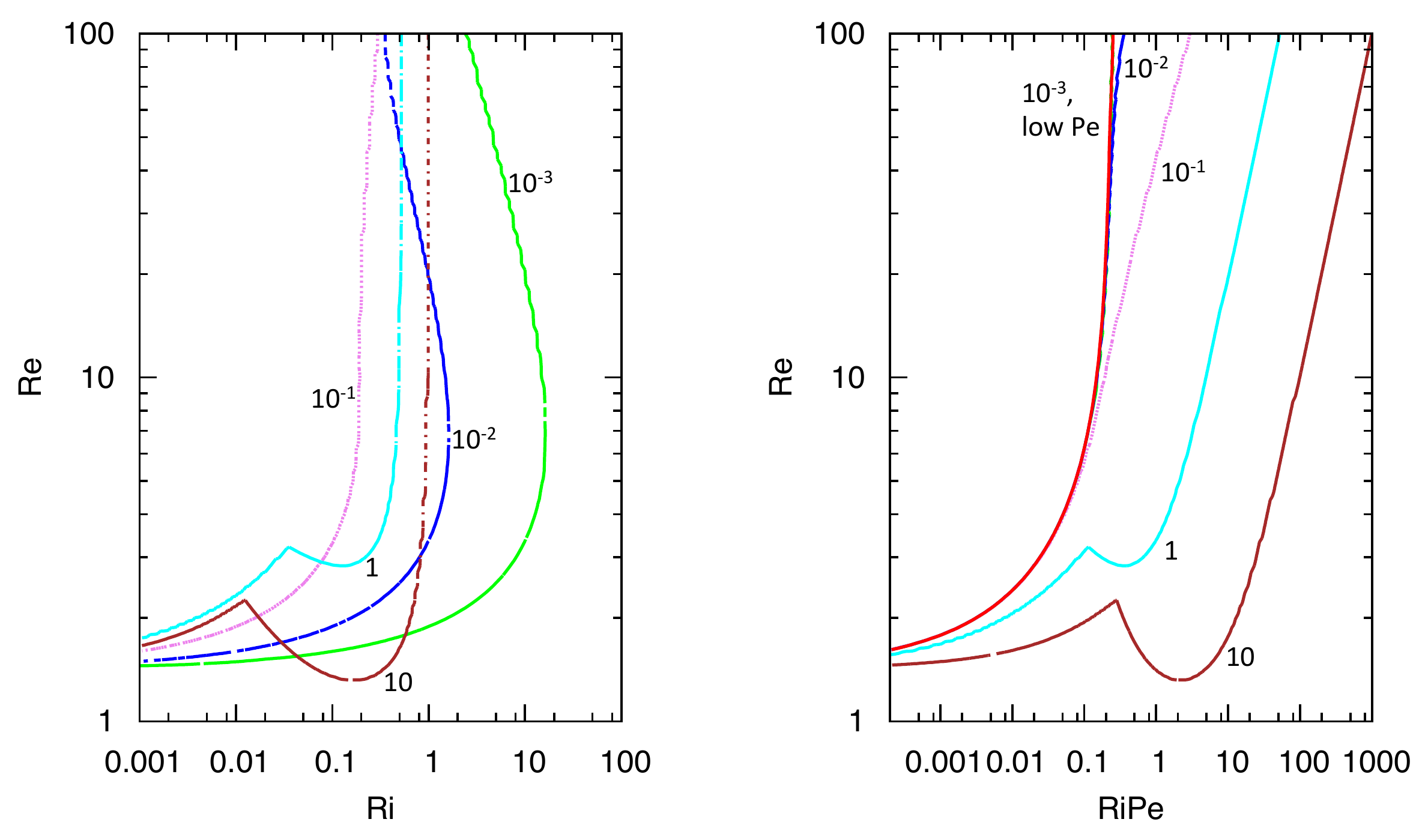}}
\caption{Left: Marginal stability curves for 2D modes in the form of $\Re$ vs. $\Ri$ for various values of the Prandtl number: $\Pr = 10^{-3}$ (green dashed line), $\Pr = 10^{-2}$ (blue small dashed line), $\Pr = 10^{-1}$ (purple dotted line), $\Pr = 1$ (cyan long dash - dotted line), and $\Pr = 10$ (brown short dash - dotted line). The system is unstable in the area above and to the left of the curves. Right: the same data plotted against $\Ri\Pe$ instead. The red solid line is the marginal stability curve for the LPN equations. The $\Pr = 10^{-3}$ and $\Pr = 10^{-2}$ curves nearly overlap with it for $\Re \le 100$, which is consistent with the fact that $\Pe < 1$ for these values of the Prandtl number. In all cases, we have truncated the Fourier expansion of $\psi'$ and $T'$ to $N=20$ to create these curves. This choice of $N$ was made after successful convergence tests.}
\label{fig:MS}
\end{figure}

Various aspects of the marginal stability surface $S_2(\Re,\Pe,\Ri) = 0$ for 2D modes are presented in Figure \ref{fig:MS}. Figure \ref{fig:MS}a shows the critical Reynolds number as a function of $\Ri$ for the standard equations, for various values of the Prandtl number ($\Pr = \Pe/\Re$). The evolution of the shape of these curves as $\Pr$ increases is not a priori easy to identify nor explain. However, an obvious result is the existence of unstable modes for reasonably large values of the Richardson number when the Prandtl number is low. This can easily be understood in the light of the work of \citet{Townsend58} (see also \citet{GageMiller74}, \citet{Jones1977},\citet{Lignieres1999},\citet{Lignieresetal1999} for instance), who showed that stratified shear instabilities can exist beyond the standard Richardson criterion when thermal diffusion is important. Since thermal diffusion increases as $\Pe$ decreases, and since $\Pe = \Pr \Re$, one can naturally expect unstable modes at high Richardson number for fixed $\Re$ and low enough $\Pr$ (or vice-versa). 

Following \citet{Lignieresetal1999}, we now show in Figure \ref{fig:MS}b the same data plotted against $\Ri\Pe$, and add the marginal stability curve for the LPN equations. The interpretation of the results is now much clearer. We first see that the LPN equations are indeed a good approximation to the full equations when the P\'eclet number is small (low $\Pr  \Re$). In the unstratified limit $\Ri \rightarrow 0$, we find that marginal stability is indeed achieved for $\Re = \sqrt{2}$, as expected \citep{Beaumont81}. We also see that, for the low-P\'eclet equations, the threshold for linear stability $(\Ri\Pe)_L$ above which the flow is linearly stable becomes independent of the Reynolds number for large enough $\Re$. The asymptotic value can be estimated numerically, and is roughly $(\Ri\Pe)_{L,\Re\rightarrow \infty} \simeq 0.25$. The fact that the critical $\Ri\Pe$ for linear stability is independent of the Reynolds number for large enough $\Re$ shows that the inviscid limit is a regular limit of this problem. This is, however, in contrast with the findings of \citet{Jones1977} and \citet{Lignieresetal1999} for the tanh shear layer. In both cases, they find that $(\Ri\Pe)_L \propto \Re$ for large $\Re$ (albeit using a fairly limited survey of parameter space). Using the LPN equations, \citet{Lignieresetal1999} also found that there is no stability threshold in the inviscid limit, that is, unstable modes exist for all values of $\Ri\Pe$. The reason for the stark difference between our results and theirs remains to be determined, but could be attributed either to the nature of the boundary conditions used (periodic vs. non-periodic), or to the fact that a sinusoidal velocity profile has shear of both signs while a tanh velocity profile only has shear of one sign. 

We now discuss linear stability to 3D perturbations using Squire's theorem. For the LPN equations, we find that the function $S_2(\Re,\Ri\Pe)$ is indeed a strictly increasing function of $\Re$, and a strictly decreasing function of $\Ri\Pe$, which implies that the marginal stability of 2D modes is also that of 3D modes (see the previous section). For larger Prandtl number, however, we can immediately see from its null contour that $S_2(\Re,\Pe,\Ri)$ is no longer a monotonic function of $\Ri\Pe$ which strongly suggests that 3D modes could be the first ones to destabilize the system. Whether this is indeed the case is beyond the scope of this paper, since it belongs to the high-P\'eclet-number regime. However, this result would be consistent with the work of \citet{SmythPeltier90}, who found that 3D modes can be the first ones to be unstable for parallel stratified shear flows which have a tanh profile, albeit in some relatively small region of parameter space. 

\begin{figure}[h]
\centerline{\includegraphics[width=0.7\textwidth]{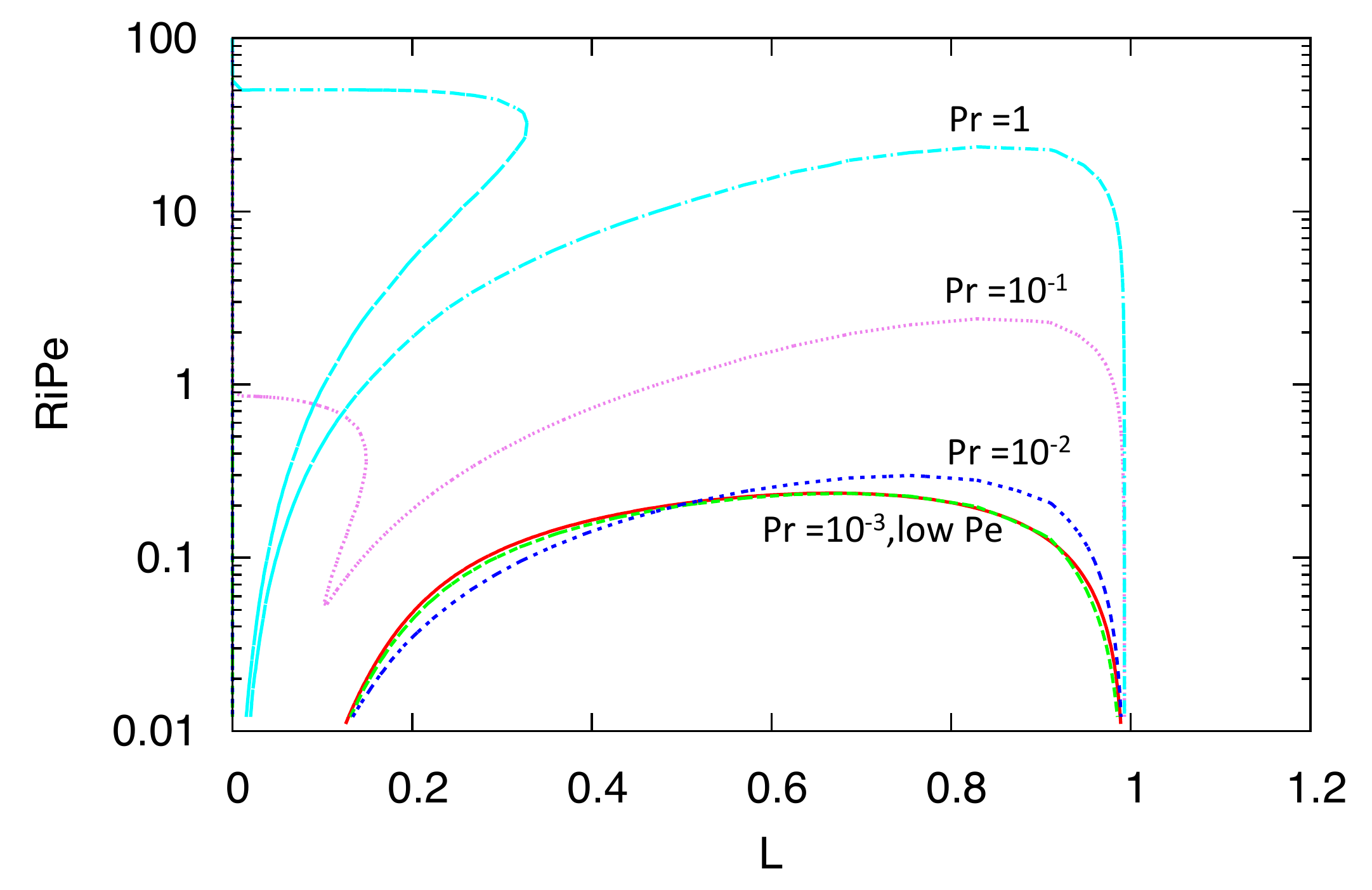}}
\caption{Marginal stability curves for 2D modes, for $\Re=10^2$, in the form of $\Ri\Pe$ vs. $L$, where $L$ is the horizontal wavenumber of the first unstable mode.
The system is unstable in the area within the curves, which therefore corresponds to the range of unstable modes for a given Richardson-P\'eclet number. The solid red line was obtained using the LPN equations. The green dashed line is for $\Pe = 0.1$ ($\Pr = 10^{-3}$), the blue short-dashed line for $\Pe = 1$ ($\Pr = 10^{-2}$), the purple dotted line for $\Pe = 10$ ($\Pr = 10^{-1}$) and the cyan dot-dashed line for $\Pe = 100$ ($\Pr = 1$). As in Figure \ref{fig:MS}, we have truncated the Fourier expansion of $\psi'$ and $T'$ to $N=20$ to create these curves.}
\label{fig:LMS}
\end{figure}

In preparation for our 3D simulations (see Section \ref{sec:num}), we are also interested in the spatial structure of the first modes to be destabilized, since the computational domain size must be chosen to be large enough to contain them in order to avoid spurious results. Based on the previous results, we now limit our study entirely to the 2D modes.  
The range of unstable 2D modes for which marginal stability is achieved is shown in Figure \ref{fig:LMS}, for both the standard equations and for the LPN equations. 
Again, we see that the results obtained using the LPN equations correctly approximate those obtained using the standard equations at low P\'eclet number. 
In all cases, we find that the first mode to be destabilized has $L \in [0,1]$, i.e. its horizontal wavelength is larger than the shear lengthscale. 
For this reason, in the numerical simulations of Section \ref{sec:num} we use a reasonably long domain size, that can fit at least two wavelengths of the most unstable mode.  

Finally, Figure \ref{fig:LMS} also sheds light on the actual source of the non-monotonicity of the 2D linear stability curves seen in Figure \ref{fig:MS} at $\Pr = 1$ and above. Indeed, it reveals a new unstable region for low horizontal wavenumber modes, which appears here for $\Pe = 10$ ($\Pr = 0.1$ in this figure). These modes have a growth rate $\lambda$ with non-zero imaginary part, by contrast with the standard shearing modes whose growth rates are real\footnote{This could be the reason why these modes were not reported by \citet{BalmforthYoung2002}..  As $\Pr$ increases, this region grows in size, and becomes the first to be destabilized for $\Pr = 1$ and above. 
However, since this new mode only exists for larger values of $\Pr$, it is less relevant for astrophysical systems. } 

\section{Energy stability}
\label{sec:ES}

\subsection{Energy stability for stratified shear flows: general ideas}

Linear stability only provides information on the stability of a shear flow to infinitesimal perturbations. Energy stability is a much stronger form of stability \citep[e.g.][]{Joseph66,Joseph76,DrazinReid}: when a system is energy stable (also called absolutely stable), perturbations of {\it arbitrarily large} amplitudes decay at least exponentially in time, and the laminar flow is the only attractor of the system. Energy stability is thus a {\it sufficient condition} for the system to be stable to perturbations of arbitrary amplitude, whereas linear instability is a {\it sufficient condition} for the system to be unstable. Often the linear stability limit and the energy stability limit do not coincide in parameter space, an extreme example being the unstratified plane Couette flow, which has a finite threshold Reynolds number for energy stability, but is linearly stable up to infinite Reynolds number: in the region between the two, the system is stable to infinitesimal perturbations, but it may be unstable to perturbations of large amplitude, i.e., it may exhibit finite-amplitude instabilities.

With the goal of further studying the stabilizing effect of background stratification, we now derive an energy stability criterion for forced stratified shear flows. 
We ask the following question: for a given amplitude of the force, is there a critical strength of the stratification above which the laminar solution is the {\it only} attractor of the system?

Again, we insert the decomposition $\bu({\bx},t) = \bu_L(z) + \bu'({\bx},t)$ in (\ref{eq:originallaminar1})-(\ref{eq:originallaminar}). However, we do not assume that $\bu'$ is small. $\bu'$ and $T'$ then satisfy
\begin{eqnarray}
&& \frac{\p \bu'}{\p t} + \bu_L \bcdot \bnabla \bu' + \bu' \bcdot \bnabla  \bu_L + \bu'\bcdot \bnabla \bu' = - \bnabla p + \Ri T' \be_z + \frac{1}{\Re} \nabla^2 \bu' \mbox{  ,} \\
&& \bnabla \bcdot  \bu' = 0 \mbox{  ,} \\
&& \frac{\p T'}{\p t} + (\bu_L + \bu') \cdot  \bnabla  T' +  w' = \frac{1}{\Pe}  \nabla^2 T'\mbox{  ,}
\end{eqnarray}
in the case of the standard equations, and 
\begin{eqnarray}
&& \frac{\p  \bu'}{\p t} + \bu_L \bcdot \bnabla  \bu' +  \bu' \bcdot \bnabla  \bu_L +  \bu' \bcdot \bnabla \bu' = - \bnabla p + \Ri\Pe \nabla^{-2} w' \be_z + \frac{1}{\Re} \nabla^2  \bu'  \mbox{  ,}  \\
&& \bnabla \bcdot  \bu' = 0 \mbox{  ,}
\end{eqnarray} 
for the LPN equations. 

An energy equation for the perturbations can be obtained by dotting the momentum equation with $\bu'$, adding it to $\gamma T'$ times the temperature equation (where the only constraint on $\gamma$ is that it should be a positive scalar), and integrating the result over the domain under consideration. Using the periodicity of the solutions, together with the incompressibility condition greatly simplifies the resulting expressions, which reduce to 
\begin{eqnarray}\label{eneq_normal}
\frac{\p}{\p t}\left[ \frac{1}{2}\langle \bu'^2 \rangle +  \frac{\gamma}{2}\langle T'^2 \rangle \right] && = (\Ri - \gamma)  \langle w'T' \rangle - \langle S_L w'u' \rangle - \frac{1}{\Re}  \langle |\bnabla \bu'|^2 \rangle - \frac{\gamma}{\Pe}   \langle |\bnabla T'|^2 \rangle \mbox{  ,} \nonumber \\
&& \equiv \mathcal{H}_\gamma \left[ \bu',T' \right] 
\end{eqnarray}
for the standard equations, where $\langle \cdot \rangle$ denotes a volume integral, and $S_L(z) = \frac{\md}{\md z} u_L(z) = \cos(z)$ denotes the local vertical shear of the laminar solution. Similarly, for the LPN equations we get
\begin{equation}\label{eneq_lowpe}
\frac{\p}{\p t} \left[ \frac{1}{2}\langle \bu'^2 \rangle \right] =  \Ri\Pe \langle w' \nabla^{-2}w' \rangle - \langle S_L w'u'  \rangle - \frac{1}{\Re}  \langle |\bnabla \bu'|^2 \rangle \equiv \mathcal{H}_{LPN}\left[ \bu' \right] \mbox{  .}
\end{equation}
 This defines the two functionals $\mathcal{H}_\gamma \left[\bu',T'\right]$ and  $\mathcal{H}_{LPN}\left[\bu'\right]$, which are both quadratic forms. The task at hand is to determine the region of parameter space $\{\Re,\Pe,\Ri\}$ or $\{\Re,\Ri\Pe\}$ where these quadratic forms are negative definite, i.e., where they are strictly negative for any possible input fields $\bu'$ (and $T'$). In this region of parameter space the system is {\it energy stable}: the right-hand-side of (\ref{eneq_normal}) or (\ref{eneq_lowpe}) is strictly negative and the perturbation decays in time, regardless of its initial amplitude. On the basis of mass and momentum conservation, the only constraints that we place on $\bu'$ is that it is divergence-free and has a vanishing average over the whole domain. 
 
Comparing $\mathcal{H}_\gamma\left[\bu',T'\right]$ and $\mathcal{H}_{LPN} \left[ \bu' \right]$, we readily see that the task of proving energy stability for stratified shear flows is more involved in the case of the full set of equations than in the case of the LPN equations. 
We therefore focus on the latter, for which we are able to obtain interesting and generic results on the stability of stratified shear flows.

\subsection{Energy Stability in the low P\'{e}clet number limit}

Focussing on the LPN equations, we first compute bounds on the location of the energy stability curve in the ($\Ri\Pe,\Re$) plane. These bounds prove useful, because they correctly describe the scaling behavior of the energy stability limit for large Reynolds number. As we shall see they also validate our numerical results, and provide a simple analytical approximation to the high Reynolds number limit. 

\subsubsection{Lower bound}

While the true energy stability boundary can only be obtained by ensuring that $H_{LPN} < 0$ for {\it all} possible perturbations $\bu'$ and $T'$, one may also ask the question of when a {\it subset} of perturbations is energy stable or unstable. If the subset is unstable, then we know that the system overall is not energy stable either. The critical $\Ri \Pe$ for energy stability for that subset is therefore a lower bound (called $(\Ri \Pe)_<$ hereafter) on the true energy stability boundary. 

We now restrict our attention to perturbations of the following form: 
\begin{eqnarray}
u' & = & B \cos({\cal K} y) \, ,  \label{u'}\\
v' & = & \frac{1}{{\cal K}} \sin({\cal K} y) \sin(z)  \, , \label{v'}\\
w' & = & \cos(z) \cos({\cal K} y) \, .\label{w'}
\end{eqnarray}
where ${\cal K}$ is the wave number of the perturbation in the $y$ direction, and $B$ is a free parameter. One can check that such perturbations are divergence-free. 

We insert (\ref{u'})-(\ref{w'}) into the quadratic form (\ref{eneq_lowpe}), recalling that $S_L=  \cos(z)$, to obtain
\begin{eqnarray}
 \frac{\mathcal{H}_{LPN}}{L_x L_y L_z} & = & - \frac{\Ri \Pe}{4({\cal K}^2+1)} - \frac{B }{4} - \frac{1}{\Re} \left[ \frac{{\cal K}^2 B^2}{2} + \frac{{\cal K}^2+1}{4} \left(1+ \frac{1}{{\cal K}^2} \right)\right]  \, .
\end{eqnarray}
This expression is a quadratic polynomial in $B$. As long as its discriminant is negative, the polynomial is negative and perturbations of the form (\ref{u'})-(\ref{w'}) decay exponentially. However, when the discriminant is positive, there will be values of $B$ corresponding to growing perturbations. The threshold for energy stability of perturbations of the form (\ref{u'})-(\ref{w'}) is therefore attained when the discriminant vanishes, which gives
\begin{eqnarray}
 (\Ri \Pe)_< & = &  \frac{{\cal K}^2+1}{8 {\cal K}^2} \Re - \frac{({\cal K}^2+1)^2\left( 1+ \frac{1}{{\cal K}^2}\right)}{\Re} \, . \label{lowerbound}
\end{eqnarray}
For large enough Reynolds number, this value is maximum for the smallest value of ${\cal K}$ that is compatible with the boundary conditions, namely ${\cal K}=2 \pi / L_y$. The corresponding lower bound on the energy stability limit of the system is 
\begin{eqnarray}
 (\Ri \Pe)_< & = &  \left[\frac{\Re L_y^2}{32 \pi^2} - \frac{\left(\frac{4 \pi^2}{L_y^2}+1 \right)\left( 1+ \frac{L_y^2}{4 \pi^2}\right)}{\Re}\right]\left(\frac{4 \pi^2}{L_y^2}+1 \right)  \, . 
\end{eqnarray}
For a given system size, the high-$\Re$ asymptotic behavior of the lower-bound is
\begin{eqnarray}
 (\Ri \Pe)_< & \simeq &  \frac{ L_y^2}{32 \pi^2}\left(\frac{4 \pi^2}{L_y^2}+1 \right)\Re  \, , \label{asymptbound}
\end{eqnarray}
which shows that the Richardson-P\'eclet number needs to be at least of the order of $\Re$ to have energy stability. This bound also indicates a strong dependence of the energy stability limit on the size of the domain. Indeed, as the transverse size $L_y$ of the domain increases, expression (\ref{asymptbound}) grows as $L_y^2$: larger domains allow for perturbations that are very weakly damped by viscosity.

The lower bound is plotted in Figure \ref{fig:EnergyStab} for a domain of size $10 \pi \times 2 \pi \times 2 \pi$, for which 
\begin{eqnarray}
 (\Ri \Pe)_< & = &  \frac{\Re}{4} - \frac{8}{\Re} \, . \label{lowerboundKolm}
\end{eqnarray}

\subsubsection{Upper bound}
\label{sec:upperbound}


Upper bounds on the energy stability limit can be obtained using rigorous estimates of the three terms in $\mathcal{H}_{LPN}$. In this subsection, we do not restrict attention to shear flows of the Kolmogorov type. Instead, we consider any smooth shear flow $u_L(z)$ along $x$ that is $2\pi$-periodic in $z$. We still use the height of the domain as the characteristic length scale, and consider a domain of size $L_x \times L_y \times 2\pi$ with periodic boundary conditions. We assume that some forcing function with amplitude $F_0$ sustains the laminar flow. The dimensionless profile has an amplitude of order unity. We prove that any such laminar flow is energy stable provided the Richardson-P\'eclet number is large enough. 

To simplify notations and avoid dealing with the inverse Laplacian operator, we introduce $\theta=T'/\Pe$, such that $w'=\nabla^2 \theta$. With these notations, the quadratic functional reads
\begin{equation}
  \mathcal{H}_{LPN}\left[ \bu' \right] =  - \Ri\Pe \langle | \bnabla \theta |^2 \rangle - \langle \frac{\md u_L}{\md z} u' \nabla^2 \theta  \rangle - \frac{1}{\Re}  \langle |\bnabla \bu'|^2 \rangle \, . \label{quadtheta}
\end{equation}

Let ${\cal T}$ be the second term of this functional. After integration by parts,
\begin{equation}
{\cal T} =  - \langle \frac{\md u_L}{\md z} u' \nabla^2 \theta  \rangle =   \langle \bnabla \left( \frac{\md u_L}{\md z} u' \right)  \cdot \bnabla \theta  \rangle =  \langle   \frac{\md u_L}{\md z}  \bnabla  u'   \cdot \bnabla \theta  \rangle + \langle     \frac{\md^2 u_L}{\md z^2}  u'  \p_z \theta  \rangle \, .
\end{equation}
Using classical inequalities, we bound this term according to
\begin{eqnarray}
\nonumber |{\cal T}| &\leq &  \sup_z  \left| \frac{\md u_L}{\md z} \right|  \langle | \bnabla  u' | \, | \bnabla \theta|  \rangle  + \sup_z  \left| \frac{\md^2 u_L}{\md z^2} \right|  \langle | u' | \, | \p_z \theta|  \rangle \\
\nonumber &\leq & \sup_z  \left| \frac{\md u_L}{\md z} \right| \sqrt{ \langle | \bnabla  u' |^2 \rangle } \sqrt{ \langle | \bnabla \theta|^2  \rangle } + \sup_z  \left| \frac{\md^2 u_L}{\md z^2} \right|  \sqrt{ \langle | u' |^2 \rangle } \sqrt{\langle | \p_z \theta|^2  \rangle } \\
\nonumber &\leq & \frac{\Ri \Pe}{2} \langle | \bnabla \theta|^2  \rangle + \frac{1}{2 \Ri \Pe} \sup_z  \left| \frac{\md u_L}{\md z} \right|^2  \langle | \bnabla  u' |^2 \rangle \\
\nonumber & & + \frac{\Ri \Pe}{2} \langle | \bnabla \theta|^2  \rangle  + \frac{1}{2 \Ri \Pe}   \sup_z  \left| \frac{\md^2 u_L}{\md z^2} \right|^2  \langle | u' |^2 \rangle \\
 & = & \Ri \Pe \langle | \bnabla \theta|^2  \rangle +  \frac{1}{2 \Ri \Pe} \sup_z  \left| \frac{\md u_L}{\md z} \right|^2  \langle | \bnabla  u' |^2 \rangle + \frac{1}{2 \Ri \Pe}   \sup_z  \left| \frac{\md^2 u_L}{\md z^2} \right|^2  \langle | u' |^2 \rangle \, , \label{ineqtemp}
\end{eqnarray}
where we have used respectively H\"older's inequality to get the first line, Cauchy-Schwartz inequality to get the second one, and Young's inequality to get the final expression (see \citet{Doering2000} for another example of the use of these inequalities in a fluid dynamics context). We now wish to express $ \langle | u' |^2 \rangle$ in terms of $\langle | \bnabla  \bu' |^2 \rangle$. To wit, we use Poincar\'e's inequality \citep[see][]{Doering2000}: the divergence-free constraint implies that $u'$ has vanishing Fourier amplitude on modes with vanishing wave numbers in both the $y$ and $z$ directions, hence
\begin{equation}
\langle |  u' |^2 \rangle \leq \frac{1}{4 \pi^2}\max \left\{ {L_y^2};  {4 \pi^2} \right\}  \langle | \bnabla  u' |^2 \rangle \, ,
\end{equation}
where the arguments of the $\max$ are the maximum allowed values for the squared wavelengths in the $y$ and $z$ directions.
Inserting this inequality into (\ref{ineqtemp}), together with $\langle | \bnabla  u' |^2 \rangle \leq \langle | \bnabla  \bu' |^2 \rangle$, leads to
\begin{eqnarray}
  \mathcal{H}_{LPN}\left[ \bu' \right] \leq    \langle |\bnabla \bu'|^2 \rangle \left[  \frac{1}{2 \Ri \Pe} \left( \sup_z  \left| \frac{\md u_L}{\md z} \right|^2  +   \sup_z  \left| \frac{\md^2 u_L}{\md z^2} \right|^2   \max \left\{ \frac{L_y^2}{4 \pi^2}; 1 \right\}  \right)        - \frac{1}{\Re} \right]   \, .\label{ineqtemp2}
\end{eqnarray}
As discussed at the beginning of this subsection, the non-dimensionalization is such that
\begin{eqnarray}
 \sup_z  \left| \frac{\md u_L}{\md z} \right|^2  & = & c_1 \, , \\
 \sup_z  \left| \frac{\md^2 u_L}{\md z^2} \right|^2 & = & c_2  \, ,
 \end{eqnarray}
where $c_1$ and $c_2$ are constants of order unity that depend on the shape of the laminar profile only (and specifically not on $\Re$, $\Ri \Pe$, $L_y$, etc). Combining these expressions with (\ref{ineqtemp2}) leads to
\begin{eqnarray}
  \mathcal{H}_{LPN}\left[ \bu' \right] \leq    \langle |\bnabla \bu'|^2 \rangle \left[  \frac{1}{2 \Ri \Pe} \left( c_1+c_2   \max \left\{ \frac{L_y^2}{4 \pi^2}; 1 \right\} \right)          - \frac{1}{\Re} \right]   \, ,\label{ineqfinal}
\end{eqnarray}
hence $\mathcal{H}_{LPN}$ is a negative quadratic form if the expression inside the square brackets is negative, i.e., if
\begin{eqnarray}
\Ri \Pe > (\Ri \Pe)_> = \left( c_1+c_2   \max \left\{ \frac{L_y^2}{4 \pi^2}; 1 \right\} \right) \, \frac{ \Re}{2} \, .\label{suffcond}
\end{eqnarray}
Because we have used rough but rigorous estimates of the different terms of the quadratic functional, $(\Ri \Pe)_>$ is an upper bound on the actual energy stability limit of the system. It proves that {\it any} smooth laminar velocity profile is absolutely stable provided the stratification is strong enough. This has important implications, showing in particular that for fixed Reynolds number and strong enough stratification such a shear flow does not induce any turbulent mixing! Note, however, that since this result is obtained for the LPN equations, it is formally only valid for perturbations that have a low P\'eclet number. This can be done mathematically by taking the asymptotic limit of the equations for $\Pe\rightarrow 0$ before considering perturbations of arbitrary amplitude. In practice, however, our result does not rule out the possibility of instability for perturbations that have a high P\'eclet number. A simple example would be perturbations that locally reduce or eliminate the horizontally-averaged vertical stratification the domain: such perturbations are not allowed in the LPN equations, but they are allowed in the full set of equations at very low P\'eclet number, where they might indeed allow for sustained turbulent solutions localized in the mixed layer. 

For domains with large extent in the $y$-direction, the sufficient condition (\ref{suffcond}) for energy stability becomes approximately $\Ri \Pe \gtrsim L_y^2 \Re$. In the particular case of the Kolmogorov velocity profile, we can compare this upper bound to the lower bound (\ref{asymptbound}): both bounds scale as $L_y^2 \Re$ for large Reynolds number, which indicates unambiguously that the actual critical $\Ri \Pe$ for energy stability obeys the same scaling. This is illustrated in Figure \ref{fig:EnergyStab}, where we plot the upper bound for a Kolmogorov flow in a domain of size $10 \pi \times 2\pi \times 2\pi$. For such a Kolmogorov flow, $c_1=c_2=1$ and the bound becomes
\begin{eqnarray}
(\Ri \Pe)_> = \Re \, .\label{upperboundKolm}
\end{eqnarray}

\subsubsection{Energy stability boundary using Euler-Lagrange equations}

To determine the true energy stability limit of the LPN system, we now consider the variational problem associated with the extremization of the quadratic functional. This gives a set of Euler-Lagrange equations, which can be solved numerically to obtain the critical value of $\Ri \Pe$ for absolute stability, called $(\Ri \Pe)_E$ hereafter. 

Starting from $\mathcal{H}_{LPN}\left[ \bu' \right]$, we separate the viscous dissipation term from the other two, as 
\begin{equation}
 \mathcal{H}_{LPN}\left[ \bu' \right] =  \Ri\Pe \langle w'\nabla^{-2} w' \rangle - \langle S_L w'u' \rangle - \mathcal{D} \mbox{  ,}
\end{equation}
where $\mathcal{D} =  \frac{1}{\Re}  \langle |\nabla \bu'|^2 \rangle$ is positive definite for non-trivial flows. We then ask the following question: for fixed viscous dissipation rate $\mathcal{D} = D_0$, for what values of $\Ri\Pe$ and $\Re$ is $\mathcal{H}_{LPN}\left[ \bu' \right] < 0$ for all incompressible velocity fields $\bu'$? As we shall see, the selected value of $D_0$ is irrelevant as it merely serves as a general normalization\footnote{One could also normalize $\bu'$ by fixing the total kinetic energy, but the advantage of setting $D_0$ is to eliminate the constant velocity solution, which is otherwise discarded on the basis of momentum conservation.} of $\bu'$.  In order to answer this question, it is sufficient to maximize $\Ri\Pe\langle w'\nabla^{-2} w' \rangle - \langle S_L w'u'  \rangle$ over all possible incompressible flows $\bu'$, and find out for what values of $\Ri\Pe $ and $\Re$ this maximum is  smaller than $D_0$. The problem thus reduces to an Euler-Lagrange optimization problem. 

Using the notation $\theta=T'/\Pe$ as in the previous section, we construct the following Lagrangian: 
\begin{equation}
\mathcal{L}\left[ \bu' \right] =  \Ri\Pe \langle  
w'\theta  \rangle - \langle S_L w'u'  \rangle + \langle \pi_1(x,y,z) \nabla \cdot \bu'\rangle - \pi_2 \langle \frac{|\nabla \bu'|^2}{\Re} - D_0 \rangle + \langle \pi_3(x,y,z) (w' -  \nabla^2 \theta ) \rangle \mbox{  ,}
\end{equation}
where the Lagrange multiplier function $\pi_1(x,y,z)$ enforces incompressibility at every point, $\pi_3(x,y,z)$ enforces equation (\ref{eq:lowPeT}) at every point, and the constant multiplier $\pi_2$ enforces $\mathcal{D} = D_0$ globally\footnote{The choice of a negative sign in front of $\pi_2$ will be clarified shortly.}. Note that we go back here to using the field $\theta$ merely to avoid dealing with inverse Laplacian operators in the variational problem; it is also possible to work through the following derivation without doing it. 

 The maximizing perturbation field $\bu'$ has to solve the Euler-Lagrange equations: three of them (arising from the derivatives of $\mathcal{L}$ with respect to $\pi_1$, $\pi_2$ and $\pi_3$) simply recover the constraints, and the other four (arising from the derivatives of $\mathcal{L}$ with respect to $u'$, $v'$, $w'$ and $\theta$) are 
\begin{eqnarray}
&& - S_L w' -\partial_x \pi_1  = - \frac{2\pi_2}{\Re} \nabla^2 u' \label{eq:EL1} \mbox{  ,} \\
&& -\partial_y \pi_1  = - \frac{2\pi_2}{\Re} \nabla^2 v' \label{eq:EL2} \mbox{  ,} \\
&& \Ri\Pe \theta + \pi_3 - S_L u'  - \partial_z \pi_1 =- \frac{2\pi_2}{\Re} \nabla^2 w' \label{eq:EL3} \mbox{  ,}\\
&& \Ri\Pe w' =  \nabla^2 \pi_3\label{eq:EL4} \mbox{  .}
\end{eqnarray}
We see that the multiplier $\pi_1$ plays a role similar to pressure, a standard result. Comparing the fourth equation with the constraint (\ref{eq:lowPeT}) also reveals that $\pi_3 = \Ri\Pe \theta $, which implies that we can eliminate both $\theta$ and $\pi_3$ to get
\begin{equation}
2\Ri\Pe \nabla^{-2} w'  - S_L u'  - \partial_z \pi_1 = -\frac{2\pi_2}{\Re} \nabla^2 w' \label{eq:ESweq} \mbox{  .}
\end{equation}
Dotting equations (\ref{eq:EL1}) to (\ref{eq:EL3}) with $\bu'$ and integrating over the domain, we get (using incompressibility) the relationship: 
\begin{equation}
-  \langle S_L u' w' \rangle + \Ri\Pe \langle  w' \nabla^{-2} w' \rangle = \frac{\pi_2}{\Re} \langle |\nabla \bu'|^2 \rangle = \pi_2 \mathcal{D} \mbox{  ,}\\
\end{equation}
which reveals the interpretation of $\pi_2$, and allows us to write 
\begin{equation}
 \mathcal{H}_{LPN}\left[ \bu' \right] = (\pi_2 - 1) \mathcal{D}  \mbox{  .}
\end{equation}
Since $\mathcal{D}$ is positive, this expression shows that energy stability corresponds to $\pi_2 < 1$.  All that is left to do is to solve equations (\ref{eq:EL1})-(\ref{eq:EL4}) as well as the incompressibility condition for the eigenvalue $\pi_2$ and determine for which values of $\Re$ and $\Ri\Pe$ the condition $\pi_2 < 1$ is satisfied.
Unfortunately, this system of equations does not generally lend itself to Squire's transformation, which means that we need to study the full 3D eigenproblem to solve for $u'$, $v'$, $w'$, $\theta$, $\pi_1$ and of course $\pi_2$.

\begin{figure}[h]
  \centerline{ \includegraphics[width=0.8\textwidth]{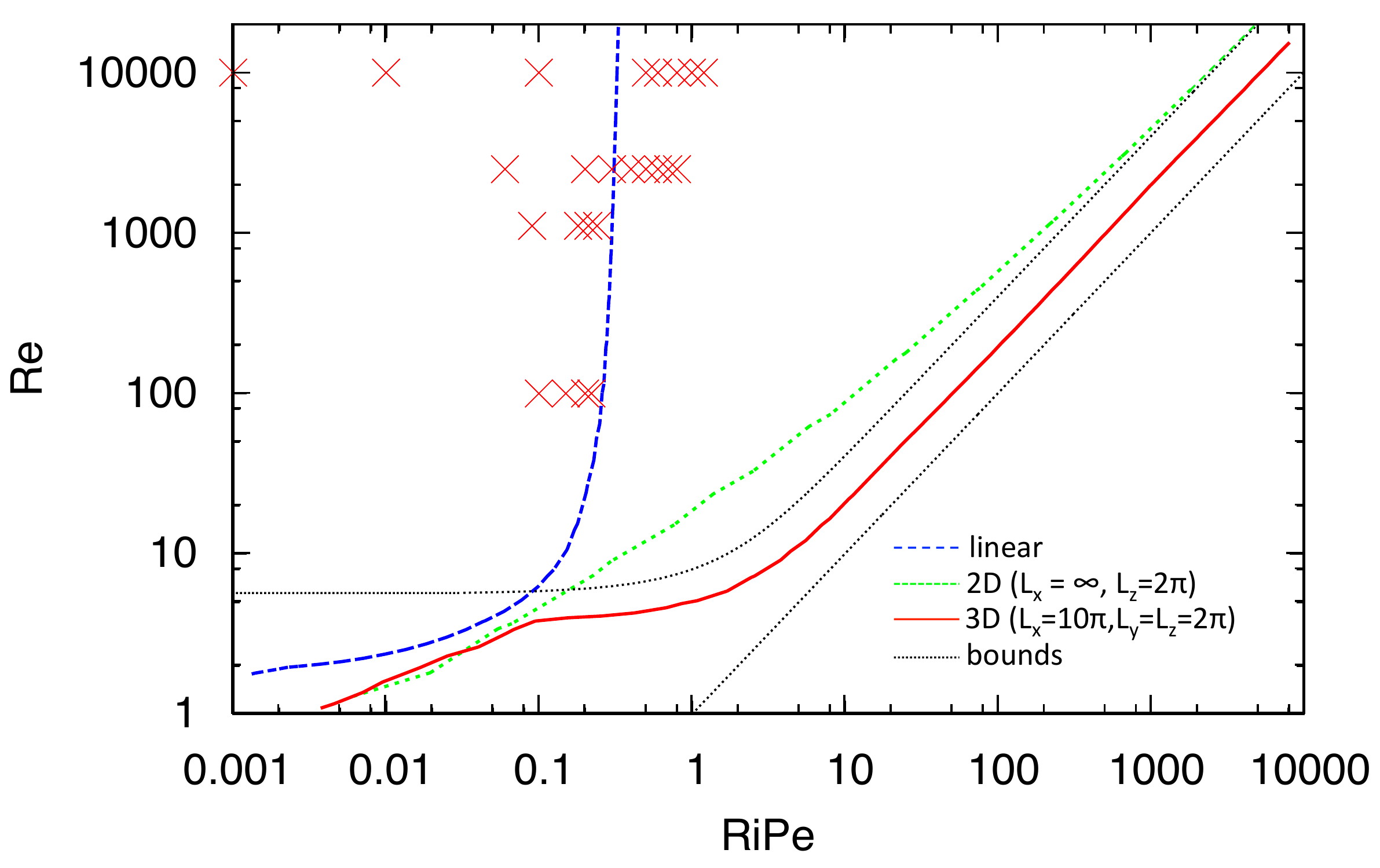}}
  \caption{Stability boundaries for the LPN equations. The linear stability boundary is valid for infinite domains in both 2D and 3D. The energy stability boundary is shown in 2D (green curve) for a domain of arbitrary horizontal extension and in 3D (red curve) for the periodic domain of size $L_x\times L_y \times L_z=10\pi \times 2\pi \times 2 \pi$ used for the low-$\Pe$ numerical simulations. The 3D energy stability limit falls between the lower and upper bounds (\ref{lowerboundKolm}) and (\ref{upperboundKolm}). At large Reynolds number, the flow is linearly stable above a critical value $(\Ri \Pe)_L$, and the energy stability limit follows the scaling $(\Ri \Pe)_E \sim \Re$. The symbols mark the simulations for which a turbulent solution was found numerically, using the LPN equations.}
\label{fig:EnergyStab}
\end{figure}


Since $S_L(z) = \cos(z)$ for the Kolmogorov flow studied here, we use Floquet theory again to solve (\ref{eq:EL1})-(\ref{eq:EL4}), together with $\pi_1  = p'$, $\pi_3 = \Ri\Pe \theta$ and the incompressibility condition. For simplicity, and for ease of comparison with the numerical simulations of the next section, we now restrict our attention to domains with vertical extent $L_z = 2\pi$ by setting the Floquet coefficient $a=0$. Assuming an ansatz of the form
\begin{equation}
q(x,y,z) =  e^{il x + i m y} \sum_{n=-N}^{N} q_n e^{inz} 
\end{equation}
for each of the unknown functions yields the system:
\begin{eqnarray}
&& - \frac{\Re}{2} (w_{n-1}+w_{n+1}) - i l  p_n   = \frac{2\pi_2}{\Re}\left( l^2 + m^2 + n^2\right)  u_n  \mbox{  ,} \\
&& -i m  p_n   = \frac{2\pi_2}{\Re} \left( l^2 + m^2 + n^2\right) v_n  \mbox{  ,} \\
&& 2 \Ri \Pe \theta_n  - \frac{\Re}{2} (u_{n-1}+u_{n+1}) - i n p_n  = \frac{2\pi_2}{\Re} \left( l^2 + m^2 + n^2\right) w_n \mbox{  ,} \\
&&  w_n  + \ \left( l^2 + m^2 + n^2\right) \theta_n = 0 \mbox{  ,} \\
&&  l u_n  + m v_n + n w_n  = 0 \mbox{  ,}
\end{eqnarray}
for $n = -N..N$. This forms a generalized eigenvalue problem $\bA \bz =\pi_2 \bB \bz $, where \\ $\bz = (u_{-N},...,u_N,v_{-N},....v_N,w_{-N},...,w_N,p_{-N},...,p_{N},\theta_{-N},...,\theta_{N})$, which can be solved numerically (using LAPACK routines) for the eigenvalue $\pi_2$. The latter depends on the horizontal wavenumbers $l$ and $m$ as well the original parameters $\Re$ and $\Ri\Pe$. For given $\Re$ and $\Ri\Pe$, energy stability is achieved if $\pi_2 < 1$ for all possible $l$ and $m$. At fixed $\Ri\Pe$, the critical Reynolds number for energy stability is the largest value of $\Re$ for which this is true. Conversely, at fixed Reynolds number, the critical Richardson-P\'eclet number for energy stability $(\Ri\Pe)_E$ is the smallest value of $\Ri\Pe$ for which $\pi_2 < 1$. The energy stability boundary therefore delimits the region of parameter space where the maximum value of $\pi_2$ over all possible $l$ and $m$ is smaller than unity.

Figure \ref{fig:EnergyStab}a compares the linear stability boundary to the energy stability boundary. The former is computed for an infinite domain and is valid both in 2D and in 3D, while the latter is computed for a 2D ($y-$independent) domain of infinite horizontal extent and for the 3D domain of size $L_x\times L_y \times L_z=10\pi \times 2\pi \times 2 \pi$ used in the low P\'eclet numerical simulations of the next section. Note how the 3D energy stability curve is lower than the 2D one, which is expected since the family of all possible 2D perturbations is included in the family of all possible 3D perturbations. Systems whose parameters lie below the 3D energy stability curve are always stable to perturbations of arbitrary amplitude, while systems whose parameters lie above the linear stability curve are unstable to infinitesimal perturbations. The 3D energy stability curve lies strictly below the linear instability curve, revealing a significant region of parameter space between them where a stratified shear flow is stable to small perturbations but could be destabilized by appropriate finite-amplitude perturbations. 

At large Reynolds number, $(\Ri \Pe)_E$ scales as $\Re$ with a proportionality constant of order unity, in agreement with the predictions from the upper and lower bounds. It is interesting to note that $(\Ri \Pe)_E \simeq \Re$ is equivalent to $(\Ri \Pr)_E \simeq O(1)$, which is reminiscent of the nonlinear stability criterion originally proposed by \citet{Zahn1974}, albeit with the right-hand-side constant of order unity rather than of order $10^{-3}$. His original argument, modified to have $\Re_{\rm crit} = 1$, could therefore provide a plausible physical explanation for the energy-stability scaling found.

In summary, we have formally proved, using both simple analytical bounding arguments and exact numerical integration of the Euler-Lagrange equations, that a strong enough stratification makes the laminar shear flow stable to perturbations of arbitrary form and amplitude, within the constraint that the perturbations must still have a low P\'eclet number (see discussion in Section \ref{sec:upperbound}).

\section{Numerical simulations}
\label{sec:num}

Our findings strongly suggest that stratified shear flows are subject to finite-amplitude instabilities, which raises the question of the relevance of linear stability analyses in determining when turbulent mixing is expected. In order to clearly assess the existence of finite-amplitude instabilities, we now turn to direct numerical simulations. 

\subsection{The numerical model}

We solve the set of equations (\ref{eq:originallaminar1})-(\ref{eq:originallaminar}) in a triply-periodic domain of size $L_x = 10\pi$, $L_y = 4\pi$ and $L_z = 2\pi$, using the pseudo-spectral code originally developed by S. Stellmach to study double-diffusive convection \citep{Traxleretal2011b,Stellmachetal2011}. This code has been modified to include the effect of the body force $\bF$. Table 1 shows a record of simulations run in this format. In all of these runs, $\Re = 10^4$, and $\Pe$ is either 0.1, 1 or 10.

We then modified the code to solve instead the LPN momentum equation (\ref{eq:lowPeeqs}) together with the continuity equation, and have run a number of simulations with this new setup in a somewhat smaller domain ($L_x = 10\pi$, $L_y = 2\pi$ and $L_z = 2\pi$), for $\Re$ ranging from $10^2$ to $10^4$. The difference in the two domain sizes does not appear to have any influence on the numerical results in the low-P\'eclet-number regime, hence our decision to save on computer time in this second set of runs. The latter are summarized in Table 2. 

\begin{table}
\begin{center}
\begin{tabular}{cccc}
\\
\tableline
$\Pe$  &$\Ri$ & $\Ri\Pe$ & Transition to turbulence \\
\tableline
0.1 & 1 & 0.1 &  Linear Instab. \\
0.1 & 10 & 1 & Fin. amp. instab. starting from $\Ri = 1$ run. \\
0.1 & 12 & 1.2 & Fin. amp. instab. starting from $\Ri = 10$ run. \\
0.1 & 15  & 1.5 & No instab. found starting from $\Ri = 12$ run. \\
\tableline
1 & 0.001 & 0.001 & Linear Instab. \\ 
1 & 0.01 & 0.01 &  Linear Instab. \\
1 & 0.1 & 0.1 & Linear Instab. \\
1 & 0.3 & 0.3 & Fin. amp. instab. starting from $\Ri = 0.1$ run. \\
1 & 0.5 & 0.5 & Fin. amp. instab. starting from $\Ri = 0.3$ run. \\
1 & 0.7 & 0.7 & Fin. amp. instab. starting from $\Ri = 0.5$ run. \\
1 & 1 & 1 & Fin. amp. instab. starting from $\Ri = 0.7$ run. \\
1 & 1.2 & 1.2 & Fin. amp. instab. starting from $\Ri = 1$ run. \\
1 & 1.5  & 1.5 & No instab. found starting from $\Ri = 1.2$ run. \\
\tableline
10 & 0.0001 & 0.001 & Linear Instab. \\ 
10 & 0.001 & 0.01 &  Linear Instab. \\
10 & 0.01 & 0.1 & Linear Instab. \\
10 & 0.1 & 1 & Fin. amp. instab. starting from $\Ri = 0.01$ run. \\
10 & 0.12 & 1.2 & Fin. amp. instab. starting from $\Ri = 0.1 $ run. \\
10 & 0.15  & 1.5 & No instab. found starting from $\Ri = 0.12 $ run. \\
\tableline
\end{tabular}
\caption{Presentation of the various runs performed using the standard equations.  All runs are at $\Re = 10^4$, in rectangular domains
  of size $10 \pi \times 4 \pi \times 2\pi$. The resolution (in terms of equivalent
  mesh-points $N_{x,y},N_z$) is the same in all directions, and for all runs, and is equal to 192 mesh points per interval of length $2\pi$. Runs that go unstable starting from infinitesimal perturbations are marked ``Linear Instab.". Runs that do not go unstable starting from infinitesimal perturbations, but that have finite-amplitude instabilities are marked ``Fin. amp. instab.". These runs were started using the endpoint of another simulation at lower $\Ri$, also noted.}
 \end{center}
\label{run-summary}
\end{table}

\begin{table}
\begin{center}
\begin{tabular}{ccc}
\\
\tableline
$\Re$ & $\Ri \Pe $ & Transition to turbulence \\
\tableline
100 & 0.1 & Linear Instab. \\
100 & 0.2 & Linear Instab. \\
100 & 0.22 & Linear Instab. \\
100 & 0.25 & No instab. found starting from $\Ri \Pe = 0.22$ \\
\tableline
1100 & 0.09 & Linear Instab. \\
1100 & 0.21 & Linear Instab. \\
1100 & 0.24 & Linear Instab. \\
1100 & 0.27 & No instab. found starting from $\Ri \Pe = 0.24$ \\
\tableline
2500 & 0.06 & Linear Instab. \\
2500 & 0.2 & Linear Instab. \\
2500 & 0.3 & Fin. amp. instab. starting from $\Ri\Pe = 0.2$ run. \\
2500 & 0.4 & Fin. amp. instab. starting from $\Ri\Pe = 0.3$ run.\\
2500 & 0.5 & Fin. amp. instab. starting from $\Ri\Pe = 0.4$ run. \\
2500 & 0.6 & Fin. amp. instab. starting from $\Ri\Pe = 0.5$ run. \\
2500 & 0.7 & Fin. amp. instab. starting from $\Ri\Pe = 0.6$ run. \\
2500 & 0.8 & Fin. amp. instab. starting from $\Ri\Pe = 0.7$ run.\\
2500 & 0.9 & No instab. found starting from  $\Ri\Pe = 0.8$ run \\
\tableline
10000 & 0.01  & Linear Instab. \\
10000 & 0.1 & Linear Instab. \\
10000 & 0.3 & Fin. amp. instab. starting from $\Ri\Pe = 0.1$ run. \\
10000 & 0.5 & Fin. amp. instab. starting from $\Ri\Pe = 0.3$ run. \\
10000 & 0.6 & Fin. amp. instab. starting from $\Ri\Pe = 0.5$ run. \\
10000 & 0.8 & Fin. amp. instab. starting from $\Ri\Pe = 0.6$ run. \\
10000 & 1 & Fin. amp. instab. starting from $\Ri\Pe = 0.8$ run. \\
10000 &1.2 & Fin. amp. instab. starting from $\Ri\Pe = 1$ run. \\
10000 & 1.5 & No instab. found starting from  $\Ri\Pe= 1.2$ run.\\
\tableline
\end{tabular}
\caption{Presentation of the various runs performed using the LPN equations. All runs are in rectangular domains
  of size $10 \pi \times 2 \pi \times 2\pi$. Those with $\Re= 10^4$ have the same effective resolution as in Table 1. Those with $\Re= 100$ have 96 meshpoints per interval of length $2\pi$ and those with $\Re = 1100$ and $2500$ have 144 meshpoints per interval of length $2\pi$.} 
 \end{center}
\label{run-summary}
\end{table}

\subsection{Typical results}
\label{sec:qualres}

\begin{figure}[h]
  \centerline{\includegraphics[width=\textwidth]{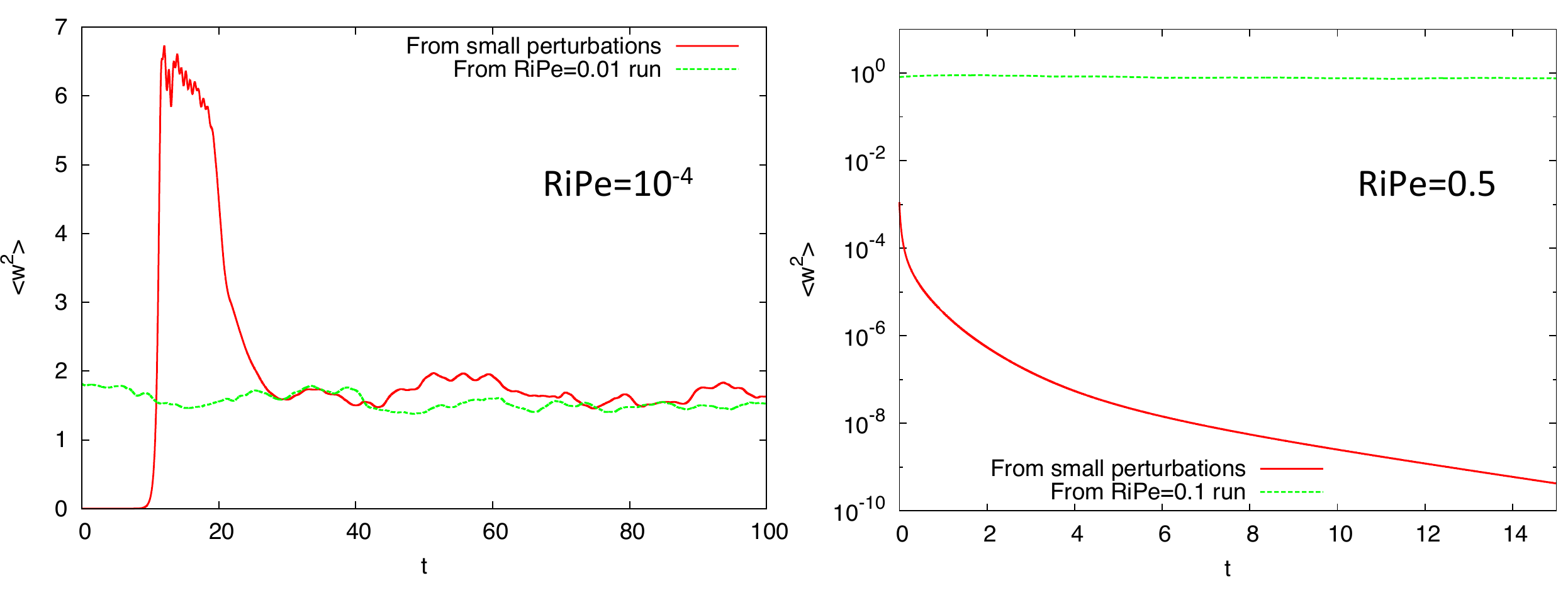}}
  \caption{Evolution of $\langle w^2 \rangle$ as a function of time, for $\Re = 10^4$ and $\Ri \Pe = 10^{-4}$ (left) and $\Ri \Pe = 0.5$ (right) starting from small random perturbations (red solid line) and from finite-amplitude perturbations (green dashed line) by continuation of a previous run at other parameter values as noted in the legend. }
\label{fig:condinit}
\end{figure}

We first take a look at typical simulations run using the LPN equations. 
\begin{figure}[h]
  \centerline{\includegraphics[width=0.8\textwidth]{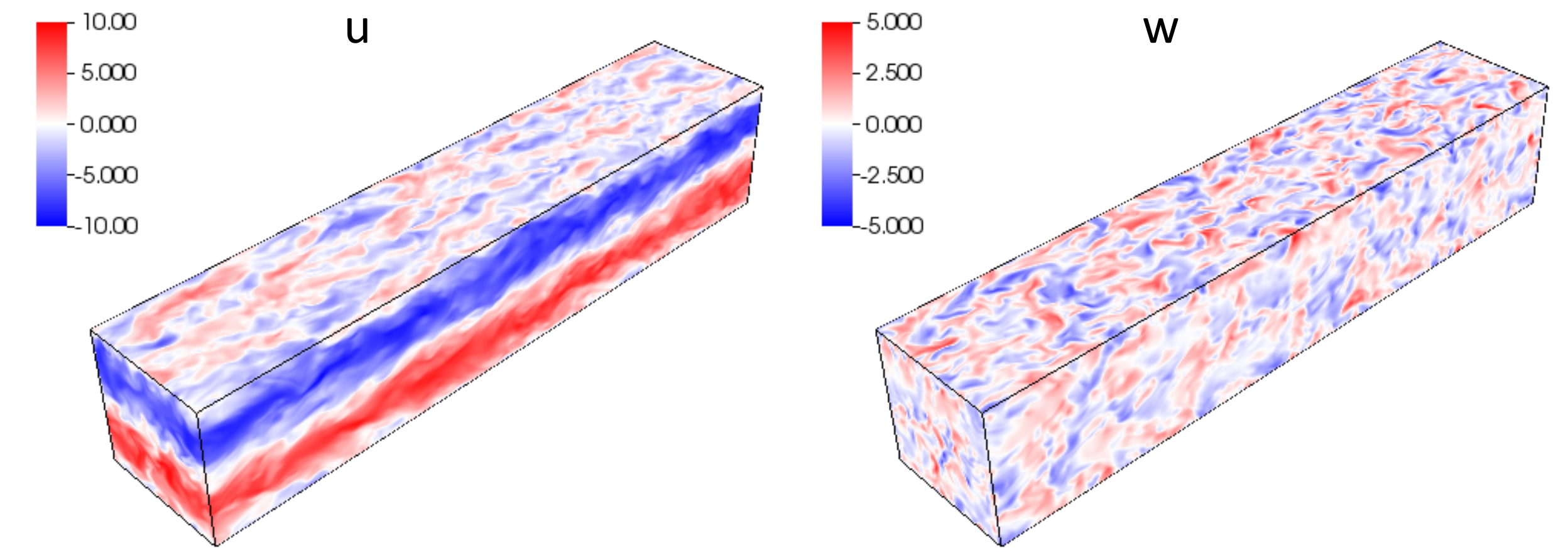}}
  \caption{Snapshot of the streamwise (left) and vertical (right) velocity components for the run at $\Re = 10^4$ and $\Ri \Pe = 0.01$ using the LPN equations, taken once it has equilibrated into a statistically-steady turbulent state.}
\label{fig:prettypic}
\end{figure}
Figure \ref{fig:prettypic} shows a system snapshot of a run at $\Re = 10^4$ and $\Ri \Pe = 0.01$, once it has equilibrated into a statistically-steady turbulent state. The shear flow is visible in the left panel (which shows the velocity field in the $x-$direction), and the typical size and amplitude of the velocity perturbations are illustrated in the right panel (which shows the velocity field in the $z-$direction). For this particular value of $\Ri \Pe$, the shear is linearly unstable. We find that whenever this is the case, the system eventually settles into a statistically-steady turbulent state that is independent of the initial conditions. This is demonstrated in Figure \ref{fig:condinit}a, which shows $\langle w^2 \rangle$ as a function of time for two simulations at $\Re = 10^4$ and $\Ri\Pe = 10^{-4}$: one that was started from small amplitude random initial conditions, and one that was started from the statistically-steady state reached by a previous run at $\Re = 10^4$ and $\Ri\Pe = 0.01$. In both cases, $\langle w^2 \rangle$ settles into the same statistically-steady state after a short transient period. The same statement applies to all global diagnostics of the system dynamics.

As shown in Figure \ref{fig:EnergyStab}a, for $\Re = 10^4$ the largest value of $\Ri \Pe$ for which the laminar steady state solution $u_L(z)$ is linearly unstable is roughly equal to $(\Ri\Pe)_L = 0.25$. We have found that all low-P\'eclet-number simulations (i.e those run using the LPN equations, and those run with the standard equations at $\Pe \le 1$) which have $\Ri \Pe < 0.25$ do indeed transition to a turbulent state, and the results shown in Figure \ref{fig:prettypic}  and Figure \ref{fig:condinit}a are fairly representative of their behavior. A detailed quantitative analysis of the results of these runs will be presented elsewhere. 

We have also found that this body-forced stratified shear flow is subject to finite-amplitude instabilities for $(\Ri \Pe)_L<\Ri \Pe<(\Ri \Pe)_c$, where the critical value $(\Ri \Pe)_c$ is discussed in the next section. This is shown in Figure \ref{fig:condinit}b, which presents $\langle w^2 \rangle$ as a function of time for two simulations at $\Re = 10^4$ and  $\Ri \Pe = 0.5$: one that was started from weak amplitude random initial conditions, and one that was started from the statistically-steady state reached by a previous run at $\Re= 10^4$ and $\Ri\Pe = 0.1$. We clearly see that the energy in the perturbations decays in the first case, but reaches a different statistically-steady state in the second case, a classical example of finite-amplitude instability.

\subsection{Finite-amplitude instability}
\label{sec:finite}

We now consider both the standard equations at $\Pe = 0.1$, $\Pe = 1$ and $\Pe = 10$ and the LPN equations. In order to find turbulent solutions for $\Ri \Pe > (\Ri \Pe)_L$ more systematically, and determine the critical value $(\Ri \Pe)_c$ for the existence of finite-amplitude instabilities, we gradually increase $\Ri$ (keeping all other parameters fixed), using as a starting solution the result of a simulation run at lower $\Ri$. We say that $(\Ri \Pe)_c$ is reached when we are no longer able to continue increasing $\Ri$ without losing the turbulent solution. Note that this only yields a rough estimate of $(\Ri\Pe)_c$ that depends largely on the size of the increments in $\Ri$ taken. It is possible that by using smaller increments one may be able to push further into the linearly stable region. Unfortunately, this is computationally very demanding and the increment size is in practice selected to satisfy our constraints on computation time. 

The results are summarized in Tables 1 and 2 and in Figure \ref{fig:EnergyStab}, and raise a number of interesting points. First note how $(\Ri \Pe)_c$ is the same for the LPN equations and for the standard equations at $\Pe = 0.1$, $\Pe = 1$ and even for $\Pe = 10$ for $\Re = 10^4$. In all cases, we have $(\Ri \Pe)_c \simeq 1.2$. This validates the use of the LPN equations as a substitute for the standard equations for low-P\'eclet-number systems. One may in fact be surprised at the fact that the LPN equations even appear to be a good approximation of the large $\Pe$ runs ($\Pe = 10$ here). However, this is due to the fact that the global P\'eclet number based on the amplitude of the hypothetical laminar shear flow $u_L$ is not a good predictor for the actual P\'eclet number of the turbulent solutions, $\Pe_{\rm rms}$ (see Section \ref{sec:lowPeeqs}). The latter is significantly smaller, and remains below one in all runs at $\Pe = 10$. This result is consistent with the theory of \citet{Lignieres1999}, which merely requires $\Pe_{\rm rms} \ll 1$ for the the LPN equations to be valid. 


The value of $(\Ri\Pe)_c \simeq 1.2$ found at $\Re = 10^4$ is somewhat larger than the linear stability threshold $(\Ri\Pe)_L$, but is significantly smaller than the theoretical energy stability limit $(\Ri\Pe)_E$ found in Section \ref{sec:ES} (see Figure \ref{fig:EnergyStab}). Some level of discrepancy is expected, since lying within the energy stability limit is only a necessary (but not sufficient) condition for instability: even though everywhere within the energy-unstable domain there exist perturbations whose amplitude initially increase with time, this does not guarantee the onset of turbulence, as in most cases transient growth is followed by rapid decay. 

These results, however, show that at $\Re = 10^4$ neither linear stability nor energy stability thresholds are good estimates for the actual threshold for transition to turbulence. Varying the Reynolds number from 100 to 10,000 we found that this is not always the case: for $\Re = 100$ and $\Re = 1100$, $(\Ri \Pe)_c$ and $(\Ri \Pe)_L$ do appear to coincide and no finite-amplitude instabilities were found. The latter only appear for $\Re = 2500$, and seem to exist at this Reynolds number for $(\Ri\Pe)_L \simeq 0.25 < \Ri \Pe <  (\Ri \Pe)_c \simeq 0.8$. 

The very limited finite-amplitude data available is not inconsistent with $(\Ri \Pe)_c \sim O(1)$ for $\Re \ge 2500$. We therefore see that, when using a non-dimensionalization based on the velocity and scale of the laminar flow, both the linear stability limit and the threshold to finite-amplitude instabilities are independent of the Reynolds number for large $\Re$ (at least, tentatively for the finite-amplitude threshold). The latter extends somewhat the stability threshold from the linear one $(\Ri\Pe)_L  \simeq 0.25$ to $(\Ri \Pe)_c \sim O(1)$, but not by a large amount. 

\section{Summary and Conclusion}
\label{sec:ccl}

We have analyzed in this work the stability of an idealized stratified, body-forced, low-P\'eclet-number shear flow using three different techniques: linear stability analysis, energy stability analysis, and direct numerical simulations. Our mathematical goal was three-fold: to test the validity of the LPN equations proposed by \citet{Lignieres1999}, to determine the respective thresholds for linear instability and energy stability, and to characterize the region of parameter space where finite-amplitude instabilities exist. 

Using dimensionless numbers based on the typical velocity of the laminar solution, our linear stability analysis confirmed that the LPN equations are indeed an excellent approximation to the standard equations of fluid dynamics provided $\Pe$ is smaller than 1. The domain of validity of these equations is in fact somewhat larger, and depends more on the P\'eclet number of the realized turbulent flow than the one of the hypothetical laminar solution. In the low P\'eclet number limit, thermal diffusion acts to destabilize the flow. We have found, as first shown by \citet{Lignieres1999} and \citet{Lignieresetal1999}, that the relevant bifurcation parameter is the Richardson number times the P\'eclet number, with stability for large Reynolds number achieved whenever $\Ri\Pe > (\Ri\Pe)_L \simeq 0.25$. This shows that shear instabilities can exist at relatively large Richardson numbers in the small P\'eclet number limit. We have also shown using an extension of Squire's transformation that in the same limit the first modes to be destabilized are 2D modes, a result which by contrast is not necessarily true for high-P\'eclet-number flows. 

We then performed an energy stability analysis of the LPN equations. We proved rigorously that any smooth low-P\'eclet-number shear flow becomes energy stable above a (Reynolds dependent) critical intensity of the background stratification. This has fundamental implications: in this region of parameter space, the laminar flow is the only attractor of the dynamics, and therefore sustained turbulent mixing cannot take place. The criterion for energy stability is approximately $\Ri \Pe \gtrsim \Re$ for large Reynolds number. Hence a laminar flow subject to strong stratification (with $\Ri \gtrsim \Pr^{-1}$) is energy stable, and the vertical diffusion of a scalar is due to molecular diffusivity only. 


These linear stability and energy stability results, however, may only be of academic interest. Indeed, using direct numerical simulations we have found that finite-amplitude instabilities in these low-P\'eclet-number stratified shear flows exist, for large enough Reynolds number, beyond the threshold for linear instabilities $(\Ri \Pe)_L \simeq 0.25$, but nevertheless disappear for $\Ri \Pe$ significantly below the threshold for energy stability $(\Ri \Pe)_E \sim \Re$. Our very limited data is consistent with a finite-amplitude instability threshold $(\Ri \Pe)_c \simeq O(1)$ for large enough $\Re$, using a non-dimensionalization based on the laminar velocity. These scaling laws are very tentative, in the sense that much remains to be done to measure $(\Ri \Pe)_c$ for larger Reynolds number, and to confirm the values found here. Indeed, as discussed above, it is possible that with more appropriately chosen initial conditions, one may be able to find turbulent solutions for even larger $\Ri \Pe$ for a given $\Re$. Furthermore, we note that while including the effects of rotation will not change the results of the energy stability analysis, it might allow for a wider range of dynamics and could help maintain turbulent solutions for larger $\Ri\Pe$. On the other hand, it is also not impossible that rotation could instead reduce the instability domain, or that some of the turbulent solutions found far into the region of linear stability are long chaotic transients that would eventually settle back to the laminar state upon longer numerical integration. 
Since the numerical constraints on the timestep and resolution increase dramatically for large $\Re$ and large $\Ri \Pe$ flows, the accurate and definitive determination of $(\Ri \Pe)_c$ at large Reynolds number is a formidable task, one that should nevertheless be undertaken in the future. 

Finally, it is also worth recalling that all of these results only apply to the low P\'eclet number regime. While sufficiently-small-scale stellar shear layers fall into that category, large-scale shear layers, however, commonly have a high P\'eclet number (albeit still with a small Prandtl number). Both linear and energy stability analyses remain to be done in this case, and may reveal further surprises.

\acknowledgements

This work was initiated as a project at the Woods Hole GFD summer program in 2013. The authors thank the program for giving them the opportunity to collaborate on this topic, and for their financial support. P. G. was also funded by NSF CAREER-0847477 and by NSF AST 1517927. B.G. was funded by the junior grant TURBA from Labex PALM ANR-10-LABX-0039. All simulations presented here were performed on the Hyades computer, purchased at UCSC with an NSF MRI grant. The authors thank S. Stellmach for providing his code. The authors also thank C. Caulfield, C. Doering, R. Kerswell, and S. Stellmach for their help and for inspiring discussions.  


\providecommand{\noopsort}[1]{}\providecommand{\singleletter}[1]{#1}%

\end{document}